\documentclass[
reprint,
 amsmath,amssymb,
 aps,
]{revtex4-2}

\usepackage{xcolor}
\usepackage{graphicx}
\usepackage{dcolumn}
\usepackage{bm}

\usepackage{amsmath,amssymb,amsthm}
\usepackage[nameinlink,capitalize,noabbrev]{cleveref}

\newtheoremstyle{ithead-plainbody}
  {6pt}{6pt}{\normalfont}{}{\itshape}{.}{0.5em}{}  

\theoremstyle{ithead-plainbody}
\newtheorem{theorem}{Theorem}

\newtheorem{corollary}{Corollary}

\newtheorem{definition}{Definition}

\begin{document}

\preprint{APS/123-QED}

\title{Exact fluctuation relation for open systems beyond the Zwanzig FEP equation}
\author{Mohammad Rahbar} 
\affiliation{Technical University of Munich; TUM School of Natural Sciences, Department of Chemistry, Lichtenbergstr. 4, D-85748 Garching, Germany}
\author{Christopher J. Stein}%
\affiliation{Technical University of Munich; TUM School of Natural Sciences, Department of Chemistry, Catalysis Research Center, Atomistic Modeling Center, Munich Data Science Institute, Lichtenbergstr. 4, D-85748 Garching, Germany}
\email{christopher.stein@tum.de}

\date{\today}

\begin{abstract}
We develop a fluctuation framework to quantify the free energy difference between two equilibrium states connected by nonequilibrium processes under arbitrary dynamics and system-environment coupling.
For an open system described by the Hamiltonian of mean force (HMF), we show that the equilibrium free energy difference between two canonical endpoints can be written as exponential averages of the HMF shift, divided by an explicit factor built from the chi-squared divergence between the initial and final system marginals.
These relations hold at the endpoint level and, under an asymptotic equilibration postulate, admit trajectory representations for general driving and coupling protocols.
A decomposition of the HMF increment along each trajectory separates the work-like contributions associated with changes in $\lambda(t)$ and $C(t)$, the heat-like exchange with the environment, and a feedback-like functional defined with respect to the initial protocol.
In the frozen-driving regime with a noninteracting reference, the equalities reduce to new FEP-like expressions involving an environment functional and an explicit overlap correction, with the Zwanzig formula recovered as a limiting case.
We validate the approach on an open system coupled to an environment and evolved under overdamped Langevin dynamics, where conventional Zwanzig FEP suffers from poor phase-space overlap and slow numerical convergence, while the present trajectory equality closely matches the exact free energy difference over a broad range of coupling strengths.
\end{abstract}

\maketitle
\section{Introduction}
\noindent
Free energy differences are key quantities in statistical mechanics and molecular simulation \cite{chipot2007free}.
They decide which chemical state is favored, how strongly molecules bind \cite{jiao2008calculation, gilson2007calculation}, when phases change \cite{hohenberg2015introduction}, and how fast reactions proceed \cite{hanggi1990reaction}.
Because of this, many simulation methods in chemistry, biology, and materials science are built around finding these differences with high accuracy \cite{cournia2024applications,york2023modern,mey2020best, duboue2021building}. 
In most cases, one does not jump directly from one state to another. Instead, one builds a path that connects two thermodynamic states, either by slowly changing interactions or by moving an external control parameter \cite{torrie1977nonphysical,chipot2007free,frenkel2023understanding,kirkwood1935statistical}.
Within this landscape, Zwanzig’s free energy perturbation theory \cite{zwanzig1954high,zwanzig2001nonequilibrium} holds a special place.
It writes the free energy difference between two canonical ensembles as an exponential average of the potential-energy difference, evaluated in the initial or final ensemble.
In this way, a simple microscopic quantity, the interaction change between two states, is turned into a macroscopic free energy difference.
This idea is both elegant and powerful, and it underpins much of modern free energy calculations.
However, the success of exponential averaging depends on subtle statistical conditions.
Good estimates require strong phase-space overlap between reference and target ensembles \cite{klimovich2015guidelines,willow2023learned}.
When the overlap is poor, the average develops a strong dependence on rare configurations \cite{chipot2007free,frenkel2023understanding}.
Then even simple-looking transformations become hard to sample, and straightforward FEP can demand very long simulations or fail to converge in practice.
These well-known issues have motivated a wide range of improvements such as enhanced sampling schemes~\cite{torrie1974monte,torrie1977nonphysical,hamelberg2004accelerated,kastner2005bridging,barducci2008well,kastner2011umbrella,plotnikov2014computing,invernizzi2020unified,swinburne2018unsupervised,falkner2024enhanced,zhang2024alchemical,wade2022alchemical,knirsch2025practical}, finely spaced intermediate states~\cite{shirts2008statistically,reinhardt2020determining}, and more advanced estimators such as overlap sampling and Bennett-type methods~\cite{bennett1976efficient,lu2004using,wu2004asymmetric,matsunaga2022use}.
All of them are, at their core, different ways of dealing with the same challenge: how to recover reliable free energy differences when the important regions of phase space are only weakly shared between the ensembles one wants to compare.
A further layer of difficulty emerges when the system of interest is strongly coupled to its environment.
In this regime, the interaction energy between system $\mathcal{S}$ and environment $\mathcal{E}$ is comparable to the bare system energy, and the standard weak coupling picture, where the bath is treated as a simple background that only sets the temperature is no longer adequate.
The Hamiltonian of mean force (HMF) is the natural tool to describe such open systems \cite{feynman2000theory,kirkwood1935statistical,roux1999implicit,talkner2016open,talkner2020colloquium,jarzynski2017stochastic,miller2018energy,seifert2016first,miller2017entropy,talkner2016open,talkner2020colloquium,anto2023effective}.
However, the HMF is difficult to construct explicitly, and its use raises conceptual issues, including ambiguities in the definition of internal energy, entropy, free energy and heat at strong coupling \cite{talkner2016open,strasberg2020measurability,campisi2009fluctuation,talkner2020colloquium,xing2024thermodynamics,colla2025observing}.
These challenges complicate the formulation of free energy methods for open systems where strong system-environment interactions are not a small correction but an essential part of the physics.
In our recent works \cite{rahbar2025thermodynamic, rahbar2025probabilistic}, we tackled this strong-coupling problem by placing the system-environment interaction energy at the center of the description.
We derived exact fluctuation equalities that split both free and internal energies into two parts: one coming from the energetic interaction itself, and one coming from how the system’s probability distribution is reshaped \cite{rahbar2025thermodynamic}.
The latter is captured by a chi-squared divergence that measures how far the coupled and uncoupled endpoint ensembles are from each other. 
The present article and companion Letter \cite{rahbar2025exact} build on this perspective and take the next step.
We ask whether one can recover the free energy difference between two equilibrium endpoints that are connected by an arbitrary nonequilibrium process, when both the underlying dynamics and the strength of system-environment coupling are  unrestricted.
In this way, we show that  equilibrium free energy perturbation and the Jarzynski equality (JE) \cite{jarzynski1997equilibrium, jarzynski1997nonequilibrium, jarzynski2004nonequilibrium,sagawa2010generalized} fit into a single strong-coupling framework and appear as special cases of a more general structure. 
We begin by formulating the evolution of the composite $\mathcal S+\mathcal E$ in terms of an abstract trajectory map $\mathcal T_t$ acting on the full phase space (Def.~\ref{def:traj}), and by specifying the total Hamiltonian along trajectories, with two independently controlled protocols $\lambda(t)$ and $C(t)$ that encode driving and coupling (Def.~\ref{def:fullH}).
On this basis, we introduce the canonical ensembles of the composite ($\mathcal S +\mathcal E$) and the marginal equilibrium distributions of the system $\mathcal S$, expressed in terms of the HMF (Def.~\ref{def:canonical}).
The same trajectory map is then used to represent the time-dependent system marginal as a pushforward of the initial equilibrium state [Eq.~\eqref{eq:traj_marginal}], which will be the key ingredient for the trajectory counterparts of our endpoint relations. 
With these kinematic and statistical ingredients in place, Thm.~\ref{thm:endpoint_equalities} establishes endpoint equalities that express the open system free energy difference $\Delta F^{*}_{\mathcal S}(\beta)$ in terms of the HMF shift $\Delta\mathcal H^{*}_{\beta}(X_{\mathcal S})$ and the chi-squared divergence $\chi^2(P^{\mathcal S}_{t_{\mathrm{eq}}}\parallel P^{\mathcal S}_{0})$ between the initial and final system marginals.
The resulting relations depend only on the canonical endpoints and remain valid for arbitrary protocols and arbitrary dynamics, as long as well-defined equilibrium states exist.
Invoking asymptotic equilibration of the full composite system, Thm.~\ref{thm:traj_endpoint} then promotes these endpoint relations to exact trajectory equalities, rewriting the same free energy difference as a trajectory-ensemble average over initial conditions $X_0\sim P_0$ propagated under arbitrary dynamics.
In this representation, the overlap factor $1+\chi^2(P^{\mathcal S}_{t_{\mathrm{eq}}}\parallel P^{\mathcal S}_{0})$ remains explicitly present, now establishing a direct link between the structure of the estimator and the statistics of trajectories underlying nonequilibrium processes via the evolved configurations $\mathcal T_{t_{\mathrm{eq}}}^{\mathcal S}(X_0)$. 
To resolve the thermodynamic content of these relations, Cor.~\ref{cor:heatwork_fixedC} derives a decomposition \cite{jarzynski2004nonequilibrium,sagawa2010generalized} of the trajectory HMF increment $\Delta\mathcal H^{*}_{\beta}(\mathcal T_{t_{\mathrm{eq}}}^{\mathcal S}(X_0))$ into three pathwise functionals: a work-like contribution $W^{*}$ driven by changes in protocols, a heat-like contribution $Q^{*}$ describing energy exchange with the environment along the actual system trajectory, and a feedback-like functional $II^{*}$ that contracts the generalized driven velocity with the generalized force field associated with the initial protocols.
Substituting this decomposition into the trajectory equalities yields exact identities in terms of $W^{*}$, $Q^{*}$, and $II^{*}$, which provide a consistent new interpretation of work, heat, and feedback-like contributions in the strong coupling regime \cite{talkner2016open, talkner2020colloquium}. 
Up to this point, the endpoint and trajectory equalities are formulated in full generality, without committing to a particular protocol.
To connect this structure to concrete free energy schemes, it is useful to reorganize it in terms of two natural protocol views.
In the frozen-coupling view, the coupling $C(t)$ is held fixed while the driving control $\lambda(t)$ carries the system between the equilibrium endpoints.
In the frozen-driving view, $\lambda(t)$ is held fixed and the transition is instead generated by switching the coupling $C(t)$.
These complementary perspectives identify which control parameter moves the system between endpoints and provide the bridge to conventional FEP and JE constructions.
The joint Letter \cite{rahbar2025exact} addresses the frozen-coupling view in full detail and its connection to the JE. 
In the remainder of the article, we focus on the frozen-driving viewpoint,
$\lambda(t)\equiv\lambda(0)$, and analyze coupling-changing protocols that
mirror the construction of FEP.
In this regime, Cors.~\ref{cor:fixedlambda_endpoint} and \ref{cor:fixedlambda}
show that the HMF shift reduces to a logarithm of a functional
$\mathcal M$, so that both the endpoint and trajectory equalities can be
written entirely in terms of ratios of $\mathcal M$ evaluated at $C(0)$ and
$C(t_{\mathrm{eq}})$, divided by the overlap factor
$1+\chi^2(P^{\mathcal S}_{t_{\mathrm{eq}}}\parallel P^{\mathcal S}_{0})$.
In Sec.~\ref{Underlying identity and FEP}, by exploiting the central identity
relating Eqs.~\eqref{eq:central_identity} and
\eqref{eq:central_free_energy_identity}, we recover the Zwanzig FEP relation
in the HMF language [Eq.~\eqref{eq:FEP_final_HMF}].
Sec.~\ref{Beyond FEP} then shows that conventional FEP is a particular
realization of the more general endpoint and trajectory equalities derived
here, and introduces a trajectory-based FEP-like estimator that contains an
explicit overlap correction.
Finally, Sec.~\ref{sec:validation_fep_like} validates this construction on an
analytically tractable system coupled to an environment
and evolved under overdamped Langevin dynamics.
There we compare the exact HMF reference, the Zwanzig estimator, and the new
trajectory equality, and we show that while conventional FEP degrades at moderate and strong coupling due to poor phase-space overlap, the proposed
trajectory relation reproduces the exact free energy difference across the entire range of coupling strengths.

\section{Preliminary definitions}
\label{Preliminary definitions}
\noindent
\begin{figure}[h]
    \centering
\includegraphics[width=\columnwidth]{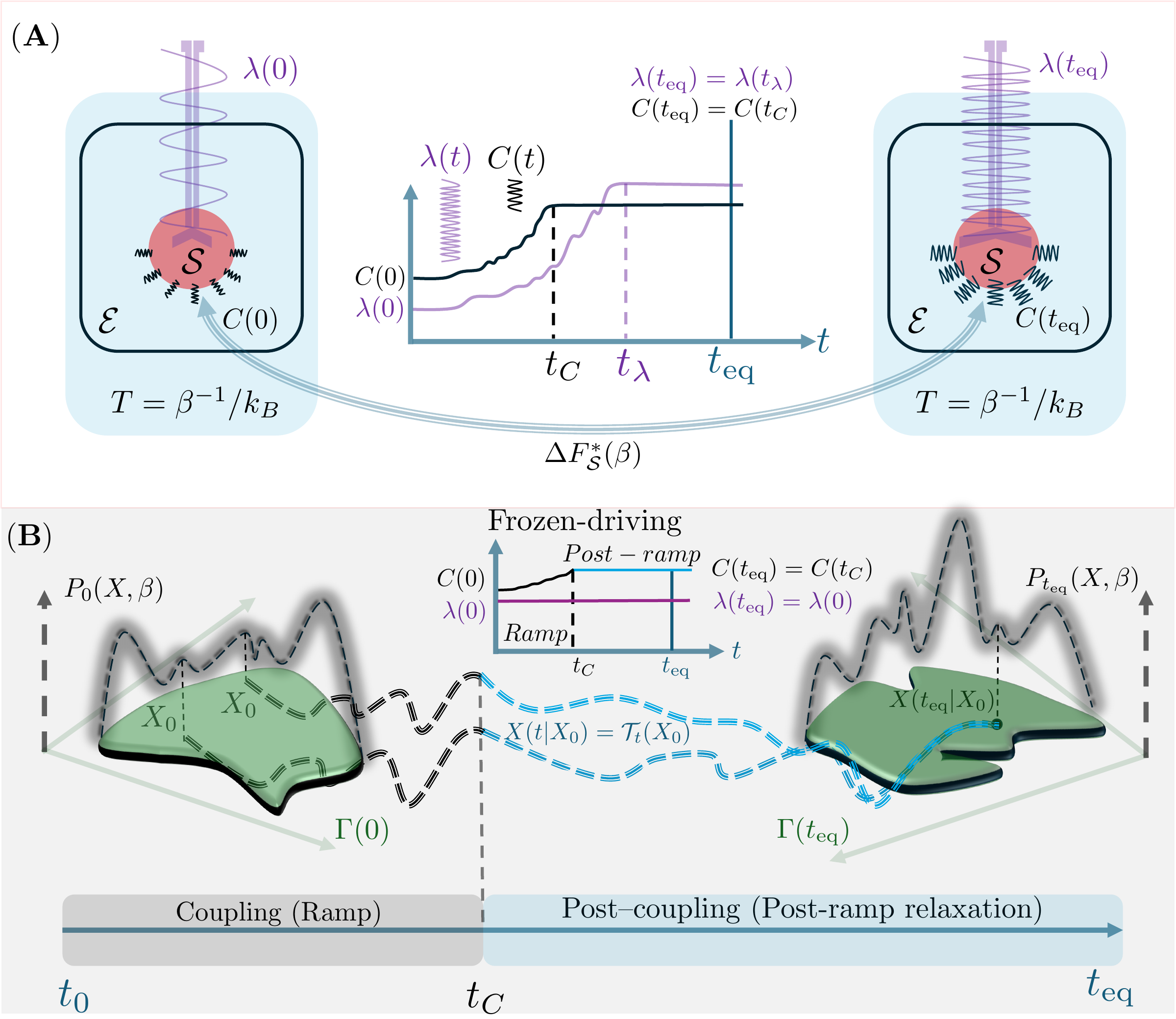}
    \caption{\textbf{(A)} Schematic of the composite $\mathcal S+\mathcal E$ at a fixed bath temperature $T=\beta^{-1}/k_B$.
The control $\lambda(t)$ (schematic purple spring) and the coupling $C(t)$ (schematic black spring) are driven independently and become constant at times $t_\lambda$ and $t_C$, respectively; the composite system then relaxes to equilibrium at $t_{\rm eq}$.
The system free energy difference $\Delta F_{\mathcal S}^*(\beta)$ refers to the canonical endpoints
$(\lambda(0),C(0))\to(\lambda(t_{\rm eq}),C(t_{\rm eq}))$ and is defined through the HMF partition functions.
\textbf{(B)} 
An initial microstate $X_0\sim P_0(X,\beta)$ on $\Gamma(0)$ evolves under the arbitrary dynamic, yielding the asymptotic density $P_{t_{\rm eq}}(X,\beta)$ on $\Gamma(t_{\rm eq})$.
The frozen driving case, $\lambda(t)\equiv \lambda(0)$, is the regime used in our validation.}
    \label{fig1}
\end{figure}
\begin{definition}[Microscopic trajectory of the composite system]
\label{def:traj}
Here, we formulate the evolution of the composite system in an abstract manner, remaining agnostic to the details of its underlying dynamics. 
We consider a composite system $\mathcal S+\mathcal E$, consisting of a system of interest $\mathcal S$ and its environment $\mathcal E$. 
The total phase space of the composite at time $t$ is denoted by $\Gamma_t$, and a single microscopic state by $X\in\Gamma_t$. 
Each microstate contains the full phase-space coordinates of both parts, $X=(X_{\mathcal S},X_{\mathcal E})$, where $X_{\mathcal S}$ and $X_{\mathcal E}$ collect the positions and momenta of particles in $\mathcal S$ and $\mathcal E$, respectively.
We introduce the trajectory map $\mathcal T_t:\Gamma_0\rightarrow\Gamma_t$, which assigns to each initial state $X_0\in\Gamma_0$ its evolved image $X(t|X_0)=\mathcal T_t(X_0)$ at time $t$. 
The map $\mathcal T_t$ is a purely kinematic construct and does not rely on any specific dynamical generator.
It may represent Hamiltonian, stochastic, or arbitrary evolution, providing a unified description for all possible dynamics. 
Since $\mathcal T_t$ acts on the composite microstates $X_0\in\Gamma_0$, no assumption is made that the system and environment trajectories can be defined independently, a requirement intrinsic to any open-system description \cite{talkner2020colloquium}. 
The microscopic state of the composite system along a trajectory is therefore written as
\begin{equation}
X(t|X_0)=
\bigl(
X_{\mathcal S}(t|X_0)=\mathcal T_t^{\mathcal S}(X_0),\,
X_{\mathcal E}(t|X_0)=\mathcal T_t^{\mathcal E}(X_0)
\bigr),
\label{eq:traj_state}
\end{equation}
where $\mathcal T_t^{\mathcal S}$ and $\mathcal T_t^{\mathcal E}$ denote the respective projections of the composite map $\mathcal T_t$ onto the system and environment coordinates. 
Next, we specify the Hamiltonian structure along the established trajectory.
\end{definition}

\begin{definition}[Total Hamiltonian along a trajectory]
\label{def:fullH}
The instantaneous total Hamiltonian of the composite system $\mathcal S+\mathcal E$ evaluated along the trajectory $\mathcal T_t(X_0)$ is defined as
\begin{align}
\label{eq:fullH}
&\nonumber
\mathcal H_{\mathcal S+\mathcal E}
\bigl(\mathcal T_t(X_0),\lambda(t),C(t)\bigr)
=
\mathcal H_{\mathcal S}
\bigl(\mathcal T_t^{\mathcal S}(X_0),\lambda(t)\bigr)
\\
&
+\mathcal H_{\mathcal E}
\bigl(\mathcal T_t^{\mathcal E}(X_0)\bigr)
+\mathcal V_{\mathcal S\mathcal E}
\bigl(
\mathcal T_t^{\mathcal S}(X_0),
\mathcal T_t^{\mathcal E}(X_0),
C(t)
\bigr).
\end{align}
Here $\mathcal H_{\mathcal S}$ and $\mathcal H_{\mathcal E}$ are the bare Hamiltonians of the system and environment, respectively, and $\mathcal V_{\mathcal S\mathcal E}$ is their interaction potential.
Two externally controlled parameters appear: $\lambda(t)$ represents the driving protocol acting on $\mathcal S$ (for example, a control parameter or mechanical coordinate), and $C(t)$ controls the coupling strength or interaction channel between $\mathcal S$ and $\mathcal E$.
By allowing $\lambda(t)$ and $C(t)$ to vary independently, we retain the most general energetic structure of an open system subject to simultaneously applied driving and coupling controls.
Each protocol may correspond to a sudden quench or to a continuous ramp in time.
We denote by $t_\lambda$ and $t_c$ the times at which the corresponding protocols become constant,
\begin{align}
\lambda(t)=\mathrm{const} &\quad\text{for } t\ge t_\lambda, 
\label{eq:tlambda}
\\
C(t)=\mathrm{const} &\quad\text{for } t\ge t_c.
\label{eq:tc}
\end{align}
The composite system subsequently relaxes to equilibrium at a later time $t_{\mathrm{eq}}$.
In general,
\begin{equation}
0\le t_\lambda,\; t_c \le t_{\mathrm{eq}}, \qquad t_\lambda\neq t_c,
\label{eq:times_relation}
\end{equation}
so the completion of driving, coupling, and equilibration need not coincide.
This distinction will later be crucial when we analyze the frozen-coupling or frozen-driving limit in which $C(t)\equiv C(0)$ or $\lambda(t)\equiv \lambda(0)$, respectively.
Now, we introduce the canonical ensembles for the composite system and the system of interest $\mathcal S$.
These equilibrium measures serve as statistical reference points for the initial and final endpoints of the process (see Fig.~\ref{fig1}).
\end{definition}

\begin{definition}[Canonical ensembles of the composite system and marginal distributions of the system]
\label{def:canonical}
The equilibrium probability distribution of the composite system $\mathcal S+\mathcal E$ at temperature $\beta^{-1}/k_{\mathrm B}$ is defined in canonical form as
\begin{equation}
P(X,\beta)
=\frac{e^{-\beta\,\mathcal H_{\mathcal S+\mathcal E}(X,\lambda,C)}}
{\mathcal Z_{\mathcal S+\mathcal E}(\lambda,C,\beta)},
\label{eq:canonical_general}
\end{equation}
where $\mathcal H_{\mathcal S+\mathcal E}$ is the total Hamiltonian given in Eq.~\eqref{eq:fullH}, and $\mathcal Z_{\mathcal S+\mathcal E}$ denotes the partition function of the composite system
\begin{equation}
\mathcal Z_{\mathcal S+\mathcal E}(\lambda,C,\beta)
=\int \mathrm dX\, e^{-\beta\,\mathcal H_{\mathcal S+\mathcal E}(X,\lambda,C)}.
\label{eq:Z_composite}
\end{equation}
Two specific equilibrium ensembles are relevant:
\begin{align}
P_0(X,\beta)
&=\frac{e^{-\beta\,\mathcal H_{\mathcal S+\mathcal E}(X,\lambda(0),C(0))}}
{\mathcal Z_{\mathcal S+\mathcal E}(\lambda(0),C(0),\beta)},
\label{eq:canonical_initial}
\\[4pt]
P_{t_{\mathrm{eq}}}(X,\beta)
&=\frac{e^{-\beta\,\mathcal H_{\mathcal S+\mathcal E}(X,\lambda(t_{\mathrm{eq}}),C(t_{\mathrm{eq}}))}}
{\mathcal Z_{\mathcal S+\mathcal E}(\lambda(t_{\mathrm{eq}}),C(t_{\mathrm{eq}}),\beta)}.
\label{eq:canonical_final}
\end{align}
The distributions $P_0$ and $P_{t_{\mathrm{eq}}}$ describe the initial and final equilibrium states of the composite system at the same bath temperature $\beta^{-1}/k_{\mathrm B}$. 
The equilibrium marginal distribution of the system $\mathcal S$ is obtained by integrating over the environmental coordinates
\begin{equation}
P^{\mathcal S}(X_{\mathcal S},\beta)
=\int \mathrm dX_{\mathcal E}\,P(X_{\mathcal S},X_{\mathcal E},\beta).
\label{eq:system_marginal}
\end{equation}
This probability can be expressed in canonical form through the HMF, defined as \cite{talkner2020colloquium,talkner2016open}
\begin{widetext}
\begin{align}
\mathcal H_{\beta}^{*}(X_{\mathcal S},\lambda,C)
=\mathcal H_{\mathcal S}(X_{\mathcal S},\lambda)
\label{eq:HMF_def}
-\frac{1}{\beta}
\ln\underbrace{
\int \mathrm dX_{\mathcal E}\,
\frac{1}{\mathcal Z_{\mathcal E}}
e^{-\beta\,[\mathcal H_{\mathcal E}(X_{\mathcal E})
+\mathcal V_{\mathcal S\mathcal E}(X_{\mathcal S},X_{\mathcal E},C)]}
}_{\mathcal M(X_\mathcal S, C, \beta)} ,
\end{align}
\end{widetext}
where 
\begin{equation}
\mathcal Z_{\mathcal E}
=\int \mathrm dX_{\mathcal E}\, e^{-\beta\,\mathcal H_{\mathcal E}(X_{\mathcal E})}.
\label{eq:Z_env}
\end{equation}
Eq.~\eqref{eq:HMF_def} defines an effective Hamiltonian that incorporates the influence of the environment on $\mathcal S$ at fixed $(\lambda,C,\beta)$.
The corresponding partition function of the system is
\begin{equation}
\mathcal Z_{\mathcal S}^{*}(\lambda,C,\beta)
=\int \mathrm dX_{\mathcal S}\,
e^{-\beta\,\mathcal H_{\beta}^{*}(X_{\mathcal S},\lambda,C)}.
\label{eq:Z_system}
\end{equation}
Hence the equilibrium marginal distribution of the system reads
\begin{equation}
P^{\mathcal S}(X_{\mathcal S},\beta)
=\frac{e^{-\beta\,\mathcal H_{\beta}^{*}(X_{\mathcal S},\lambda,C)}}
{\mathcal Z_{\mathcal S}^{*}(\lambda,C,\beta)}.
\label{eq:system_canonical}
\end{equation}
Applying this to the two equilibrium endpoints yields
\begin{align}
P^{\mathcal S}_0(X_{\mathcal S},\beta)
&=\frac{e^{-\beta\,\mathcal H_{\beta}^{*}(X_{\mathcal S},\lambda(0),C(0))}}
{\mathcal Z_{\mathcal S}^{*}(\lambda(0),C(0),\beta)},
\label{eq:HMF_initial}
\\[4pt]
P^{\mathcal S}_{t_{\mathrm{eq}}}(X_{\mathcal S},\beta)
&=\frac{e^{-\beta\,\mathcal H_{\beta}^{*}(X_{\mathcal S},\lambda(t_{\mathrm{eq}}),C(t_{\mathrm{eq}}))}}
{\mathcal Z_{\mathcal S}^{*}(\lambda(t_{\mathrm{eq}}),C(t_{\mathrm{eq}}),\beta)}.
\label{eq:HMF_final}
\end{align}
Eqs.~\eqref{eq:HMF_initial} and \eqref{eq:HMF_final} define the initial and final equilibrium probability distributions of the system in terms of its HMF, corresponding to the protocol endpoints $(\lambda(0),C(0))$ and $(\lambda(t_{\mathrm{eq}}),C(t_{\mathrm{eq}}))$.
In Thm.~\ref{thm:endpoint_equalities}, we show the exact connection between the system free energy difference at two endpoints and the probability distributions, together with their associated HMFs.
\end{definition}

\begin{definition}[Probability along  the trajectory]
Using the trajectory map $\mathcal T_t:\Gamma_0\rightarrow\Gamma_t$ defined in Def.~\ref{def:traj}, the time evolution of the probability density of the composite system is
\begin{equation}
P_t(X,\beta)
=\int_{\Gamma_0}\mathrm dX_0\,
\delta\bigl(X-\mathcal T_t(X_0)\bigr)\,P_0(X_0,\beta),
\label{eq:full_pushforward}
\end{equation}
where $P_0(X_0,\beta)$ is the initial canonical distribution given in Eq.~\eqref{eq:canonical_initial}. 
Integrating over the environmental degrees of freedom yields the evolution of the system marginal
\begin{equation}
P_t^{\mathcal S}(X_{\mathcal S},\beta)
=\int_{\Gamma_0}\mathrm dX_0\,P_0(X_0,\beta)\,
\delta\bigl(X_{\mathcal S}-\mathcal T_t^{\mathcal S}(X_0)\bigr).
\label{eq:traj_marginal}
\end{equation}

\noindent
\\
\noindent
\textit{Proof of Eq.~\eqref{eq:traj_marginal}.}  
Starting from Eq.~\eqref{eq:full_pushforward},
\begin{widetext}
\begin{equation}
P_t(X_{\mathcal S},X_{\mathcal E},\beta)
=\int_{\Gamma_0} \mathrm dX_0\;
\delta\bigl((X_{\mathcal S},X_{\mathcal E})-\mathcal T_t(X_0)\bigr)\,P_0(X_0,\beta),
\end{equation}    
\end{widetext}
with $X=(X_{\mathcal S},X_{\mathcal E})$, the system marginal follows by integrating over $X_{\mathcal E}$.  
Exchanging the order of integration and applying the property of the Dirac-delta function gives
\begin{widetext}
\begin{align}
P_t^{\mathcal S}(X_{\mathcal S},\beta)
&=\int\mathrm dX_{\mathcal E}\,P_t(X_{\mathcal S},X_{\mathcal E},\beta)
=\int\mathrm dX_{\mathcal E}\int_{\Gamma_0}\mathrm dX_0\,
\underbrace{\delta\bigl(X_{\mathcal S}-\mathcal T_t^{\mathcal S}(X_0)\bigr)
\delta\bigl(X_{\mathcal E}-\mathcal T_t^{\mathcal E}(X_0)\bigr)}_{\delta\bigl((X_{\mathcal S},X_{\mathcal E})-\mathcal T_t(X_0)\bigr)}\,P_0(X_0,\beta)
\nonumber\\
&=\int_{\Gamma_0}\mathrm dX_0\,P_0(X_0,\beta)\,\delta\bigl(X_{\mathcal S}-\mathcal T_t^{\mathcal S}(X_0)\bigr)
\underbrace{\int\mathrm dX_{\mathcal E}\,
\delta\bigl(X_{\mathcal E}-\mathcal T_t^{\mathcal E}(X_0)\bigr)}_{1}
=\int_{\Gamma_0}\mathrm dX_0\,P_0(X_0,\beta)\,
\delta\bigl(X_{\mathcal S}-\mathcal T_t^{\mathcal S}(X_0)\bigr),
\end{align}
\end{widetext}
yielding Eq.~\eqref{eq:traj_marginal}.
\hfill$\square$
\end{definition}
\section{Endpoint equalities and their trajectory counterparts}
\label{Endpoint equalities and their trajectory counterparts}
\noindent
\begin{theorem}[Endpoint equalities for free energy differences]
\label{thm:endpoint_equalities}
For two equilibrium endpoints of the composite system $\mathcal S+\mathcal E$ prepared at the same temperature $\beta^{-1}/k_B$, we have 
\begin{equation}
e^{-\beta \Delta F^{*}_{\mathcal S}(\beta)}
=
\frac{
\bigl\langle
e^{-\beta \Delta \mathcal H^{*}_{\beta}(X_{\mathcal S})}
\bigr\rangle_{\mathcal S}
}{
1+\chi^2\bigl(P^{\mathcal S}_{t_{\mathrm{eq}}}\parallel P^{\mathcal S}_{0}\bigr)
},
\label{eq:endpoint_ratio}
\end{equation}
and
\begin{equation}
e^{+\beta \Delta F^{*}_{\mathcal S}(\beta)}
=
\bigl\langle
e^{+\beta \Delta \mathcal H^{*}_{\beta}(X_{\mathcal S})}
\bigr\rangle_{\mathcal S}.
\label{eq:endpoint_moment}
\end{equation}
The free energy difference is
\begin{equation}
\Delta F^{*}_{\mathcal S}(\beta)
=
F^{*}_{\mathcal S}(\lambda(t_{\mathrm{eq}}),C(t_{\mathrm{eq}}),\beta)
-
F^{*}_{\mathcal S}(\lambda(0),C(0),\beta),
\label{eq:DeltaF_def}
\end{equation}
where $F^{*}_{\mathcal S}(\lambda,C,\beta)=-\beta^{-1}\ln \mathcal Z_{\mathcal S}^{*}(\lambda,C,\beta)$, and the HMF shift is
\begin{align}
\Delta \mathcal H^{*}_{\beta}(X_{\mathcal S})
&=
\mathcal H^{*}_{\beta}(X_{\mathcal S},\lambda(t_{\mathrm{eq}}),C(t_{\mathrm{eq}}))
\label{eq:DeltaH_def}
\\&
-
\mathcal H^{*}_{\beta}(X_{\mathcal S},\lambda(0),C(0)).
\nonumber
\end{align}
Averages are taken over the final equilibrium ensemble,
\begin{equation}
\langle\bullet\rangle_{\mathcal S}
=
\int \mathrm dX_{\mathcal S}\,
\bullet\,P^{\mathcal S}_{t_{\mathrm{eq}}}(X_{\mathcal S},\beta),
\label{eq:avg_S}
\end{equation}
and the chi-squared divergence between the endpoint marginals is
\begin{equation}
1+\chi^2\bigl(P^{\mathcal S}_{t_{\mathrm{eq}}}\parallel P^{\mathcal S}_{0}\bigr)
=
\int \mathrm dX_{\mathcal S}\,
\frac{\bigl(P^{\mathcal S}_{t_{\mathrm{eq}}}(X_{\mathcal S},\beta)\bigr)^2}
{P^{\mathcal S}_{0}(X_{\mathcal S},\beta)}.
\label{eq:chi2_def}
\end{equation}
\end{theorem}

\noindent
The result established here coincides with the endpoint equality derived in \cite{rahbar2025thermodynamic}, where the protocol is implicitly present.
As seen from the derivation of Eqs.~\eqref{eq:endpoint_ratio} and \eqref{eq:endpoint_moment}, no restriction is imposed on either the coupling or the form of the dynamics.
The reasoning relies only on the existence of well-defined equilibrium endpoints, which guarantees that the corresponding marginal distributions and free energy difference are statistically and thermodynamically meaningful.
\noindent
\\
\noindent
\textit{Proof of Thm.~\ref{thm:endpoint_equalities}}.  
From Eqs.~\eqref{eq:HMF_initial}, \eqref{eq:HMF_final}, and \eqref{eq:DeltaH_def}, we have
\begin{align}
e^{-\beta \Delta \mathcal H^{*}_{\beta}(X_{\mathcal S})}
=\frac{
P^{\mathcal S}_{t_{\mathrm{eq}}}(X_{\mathcal S},\beta)\,
\mathcal Z_{\mathcal S}^{*}(\lambda(t_{\mathrm{eq}}),C(t_{\mathrm{eq}}),\beta)
}{
P^{\mathcal S}_0(X_{\mathcal S},\beta)\,
\mathcal Z_{\mathcal S}^{*}(\lambda(0),C(0),\beta)
},
\label{eq:app_key_minus}
\end{align}
and averaging Eq.~\eqref{eq:app_key_minus} over the final ensemble, Eq.~\eqref{eq:avg_S}, yields
\begin{align}
\Bigl\langle e^{-\beta \Delta \mathcal  H^{*}_{\beta}}\Bigr\rangle_{\mathcal S}
&\nonumber=
\frac{
\mathcal Z_{\mathcal S}^{*}(\lambda(t_{\mathrm{eq}}),C(t_{\mathrm{eq}}),\beta)
}{
\mathcal Z_{\mathcal S}^{*}(\lambda(0),C(0),\beta)
}
\label{eq:app_avg_minus}
\\
&\times
\int \mathrm dX_{\mathcal S}\,
\frac{
\bigl(P^{\mathcal S}_{t_{\mathrm{eq}}}(X_{\mathcal S},\beta)\bigr)^2
}{
P^{\mathcal S}_0(X_{\mathcal S},\beta)
}.
\end{align}
Using Eq.~\eqref{eq:DeltaF_def} for the partition function ratio and Eq.~\eqref{eq:chi2_def} for the divergence, Eq.~\eqref{eq:app_avg_minus} reduces to
\begin{equation}
\Bigl\langle e^{-\beta \Delta \mathcal H^{*}_{\beta}}\Bigr\rangle_{\mathcal S}
=
e^{-\beta \Delta F^{*}_{\mathcal S}(\beta)}
\Bigl[1+\chi^2\bigl(P^{\mathcal S}_{t_{\mathrm{eq}}}\parallel P^{\mathcal S}_{0}\bigr)\Bigr],
\label{eq:app_endpoint_ratio}
\end{equation}
which rearranges to Eq.~\eqref{eq:endpoint_ratio}.
For the positive exponential, Eqs.~\eqref{eq:HMF_initial}, \eqref{eq:HMF_final}, and \eqref{eq:DeltaH_def} give
\begin{equation}
e^{+\beta \Delta \mathcal H^{*}_{\beta}(X_{\mathcal S})}
=\frac{
P^{\mathcal S}_0(X_{\mathcal S},\beta)\,
\mathcal Z_{\mathcal S}^{*}(\lambda(0),C(0),\beta)
}{
P^{\mathcal S}_{t_{\mathrm{eq}}}(X_{\mathcal S},\beta)\,
\mathcal Z_{\mathcal S}^{*}(\lambda(t_{\mathrm{eq}}),C(t_{\mathrm{eq}}),\beta)
},
\label{eq:app_key_plus}
\end{equation}
and averaging Eq.~\eqref{eq:app_key_plus} with Eq.~\eqref{eq:avg_S} cancels the denominator and yields
\begin{align}
\Bigl\langle e^{+\beta \Delta \mathcal H^{*}_{\beta}}\Bigr\rangle_{\mathcal S}
=
\frac{
\mathcal Z_{\mathcal S}^{*}(\lambda(0),C(0),\beta)
}{
\mathcal Z_{\mathcal S}^{*}(\lambda(t_{\mathrm{eq}}),C(t_{\mathrm{eq}}),\beta)
}
=
e^{+\beta \Delta F^{*}_{\mathcal S}(\beta)},
\label{eq:app_endpoint_moment}
\end{align}
reproducing Eq.~\eqref{eq:endpoint_moment}.
\hfill$\square$
\noindent
\\
\noindent
Before introducing the trajectory counterparts, it is essential to clarify the assumption underlying their derivation.
We postulate \emph{asymptotic equilibration} of the full probability distribution along the trajectory.
This assumption expresses the natural requirement that the Helmholtz free energy difference $\Delta F^{*}_{\mathcal S}(\beta)$ be well defined from the trajectory perspective of any nonequilibrium process connecting two equilibrium states, without imposing any particular dynamical constraint during the application of the protocol (i.e.\ over $[0,t_\lambda]$ and $[0,t_C]$). 

\begin{theorem}
\label{thm:traj_endpoint}
We postulate asymptotic equilibration, which implies
\begin{equation}
\lim_{t\rightarrow t_{\mathrm{eq}}}P_t^{\mathcal S}(X_{\mathcal S},\beta)
=
P^{\mathcal S}_{t_{\mathrm{eq}}}(X_{\mathcal S},\beta).
\label{eq:system_relax}
\end{equation}
Under this condition, the ensemble averages in Thm.~\ref{thm:endpoint_equalities} admit exact trajectory representations, where
\begin{equation}
\langle\bullet\rangle_{X_0}
=\int \mathrm dX_0\,\bullet\,P_0(X_0,\beta).
\end{equation}
Thus, the endpoint equalities \eqref{eq:endpoint_ratio} and \eqref{eq:endpoint_moment} take the trajectory form
\begin{equation}
e^{-\beta \Delta F^{*}_{\mathcal S}(\beta)}
=
\frac{
\Bigl\langle
e^{-\beta\Delta\mathcal H^{*}_{\beta}
\bigl(\mathcal T_{t_{\mathrm{eq}}}^{\mathcal S}(X_0)\bigr)}
\Bigr\rangle_{X_0}
}{
1+\chi^2\bigl(P^{\mathcal S}_{t_{\mathrm{eq}}}\parallel P^{\mathcal S}_{0}\bigr)
},
\label{eq:traj_ratio}
\end{equation}
and
\begin{equation}
e^{+\beta \Delta F^{*}_{\mathcal S}(\beta)}
=
\Bigl\langle
e^{+\beta\Delta\mathcal H^{*}_{\beta}
\bigl(\mathcal T_{t_{\mathrm{eq}}}^{\mathcal S}(X_0)\bigr)}
\Bigr\rangle_{X_0}.
\label{eq:traj_moment}
\end{equation}
where
\begin{align}
\Delta\mathcal H^{*}_{\beta}
\bigl(\mathcal T_{t_{\mathrm{eq}}}^{\mathcal S}(X_0)\bigr)
&\nonumber=
\mathcal H_{\beta}^*\bigl(\mathcal T_{t_{\rm eq}}^{\mathcal S}(X_0),\lambda(t_{\rm eq}),C(t_{\rm eq})\bigr)
\label{eq:decom_traj}
\\&-
\mathcal H_{\beta}^*\bigl(\mathcal T_{t_{\rm eq}}^{\mathcal S}(X_0),\lambda(0),C(0)\bigr).
\end{align}
\end{theorem}

\noindent
\textit{Proof of the numerator in Eq.~\eqref{eq:traj_ratio}.}  
Starting from the numerator of Eq.~\eqref{eq:endpoint_ratio} and using Eqs.~\eqref{eq:avg_S}, \eqref{eq:system_relax}, and \eqref{eq:traj_marginal},  
\begin{widetext}
\begin{align}
\label{eq:num-time-t}
\Big\langle e^{-\beta \Delta\mathcal H^*_\beta(X_{\mathcal S})}\Big\rangle_{\mathcal S}
&\nonumber=
\int \mathrm dX_{\mathcal S}\;
e^{-\beta \Delta\mathcal H^*_\beta(X_{\mathcal S})}\,
P_{t_{\rm eq}}^{\mathcal S}(X_{\mathcal S},\beta) =
\lim_{t\to t_{\rm eq}}
\int \mathrm dX_{\mathcal S}\;
e^{-\beta \Delta\mathcal H^*_\beta(X_{\mathcal S})}\,
P_{t}^{\mathcal S}(X_{\mathcal S},\beta)
\\
&\nonumber= \lim_{t\to t_{\rm eq}}
\int \mathrm dX_{\mathcal S}\;
e^{-\beta \Delta\mathcal H^*_\beta(X_{\mathcal S})}\,
\int_{\Gamma_0}\mathrm dX_0\,P_0(X_0,\beta)\,
\delta\bigl(X_{\mathcal S}-\mathcal T_t^{\mathcal S}(X_0)\bigr)
\\&\nonumber= \int_{\Gamma_0}\mathrm dX_0\,P_0(X_0,\beta)\,\lim_{t\to t_{\rm eq}}
\underbrace{\int \mathrm dX_{\mathcal S}\;
e^{-\beta \Delta\mathcal H^*_\beta(X_{\mathcal S})}\,\delta\bigl(X_{\mathcal S}-\mathcal T_t^{\mathcal S}(X_0)\bigr)}_{e^{-\beta \Delta\mathcal H^*_\beta
(\mathcal T_{t}^{\mathcal S}(X_0))}}\\&=\int_{\Gamma_0}\mathrm dX_0\,
P_0(X_0,\beta)\,e^{-\beta \Delta\mathcal H^*_\beta
\bigl(\mathcal T_{t_{\rm eq}}^{\mathcal S}(X_0)\bigr)}\,
=
\Big\langle
e^{-\beta \Delta\mathcal H^*_\beta
\bigl(\mathcal T_{t_{\rm eq}}^{\mathcal S}(X_0)\bigr)}
\Big\rangle_{X_0}.
\end{align} 
\end{widetext}
This confirms the numerator in Eq.~\eqref{eq:traj_ratio}.  
Applying the same steps yields  Eq.~\eqref{eq:traj_moment}.
\hfill$\square$
\section{Thermodynamic structure}
\label{Thermodynamic structure}
\begin{corollary}[Heat-work-feedback-like decomposition]
\label{cor:heatwork_fixedC}
We now discuss the thermodynamic structure encoded in Eqs.~\eqref{eq:traj_ratio} and \eqref{eq:traj_moment}.
In regimes where both driving and coupling are activated, energetic changes in $\mathcal S$ arise from (i) the manipulations of ($\lambda(t)$, $C(t)$) and (ii) exchange with the environment through the microscopic evolution of $X_{\mathcal E}$.
To inspect these contributions, we start from Eq.~\eqref{eq:decom_traj} and apply a simple algebraic insertion-subtraction of the HMF at fixed protocols, $\mathcal H^{*}_{\beta}\bigl(X_{\mathcal S}(0|X_0),\lambda(0),C(0)\bigr)$, which yields the following representation.
\begin{widetext}
\begin{align}
\label{eq:split_diff}
\Delta\mathcal H^{*}_{\beta}
\bigl(\mathcal T_{t_{\mathrm{eq}}}^{\mathcal S}(X_0)\bigr)
&=
\underbrace{\Big[
\mathcal H^{*}_{\beta}\bigl(X_{\mathcal S}(t_{\mathrm{eq}}|X_0),\lambda(t_{\mathrm{eq}}),C(t_{\mathrm{eq}})\bigr)
-\mathcal H^{*}_{\beta}\bigl(X_{\mathcal S}(0|X_0),\lambda(0),C(0)\bigr)\Big]}_{I^*(t_{\mathrm{eq}}|X_0)}
\nonumber\\&
-\underbrace{\Big[
\mathcal H^{*}_{\beta}\bigl(X_{\mathcal S}(t_{\mathrm{eq}}|X_0),\lambda(0),C(0)\bigr)
-\mathcal H^{*}_{\beta}\bigl(X_{\mathcal S}(0|X_0),\lambda(0),C(0)\bigr)\Big]}_{II^*(t_{\mathrm{eq}}|X_0)}.
\end{align}  
\end{widetext}

\noindent
\textit{Term $I^*$.}
Along each realization, the system follows the trajectory 
$X_{\mathcal S}(t|X_0)$ 
in its phase space. 
The total time variation of the HMF,
$\mathcal H^{*}_{\beta}\bigl(X_{\mathcal S}(t|X_0),\lambda(t),C(t)\bigr)$,
is obtained by applying the chain rule,
\begin{widetext}
\begin{align}
\label{eq:dH_dt}
\frac{\mathrm d}{\mathrm dt}\,
\mathcal H^{*}_{\beta}\bigl(X_{\mathcal S}(t|X_0),\lambda(t),C(t)\bigr)
=
\underbrace{\bigl(\nabla_{X_{\mathcal S}}\mathcal H^{*}_{\beta}\bigr)\cdot\dot X_{\mathcal S}}_{\dot Q^{*}}
+
\underbrace{\frac{\partial \mathcal H^{*}_{\beta}}{\partial\lambda}\,\dot\lambda
+\frac{\partial \mathcal H^{*}_{\beta}}{\partial C}\,\dot C}_{\dot W^{*}} .
\end{align}
\end{widetext}
The first term describes the energy exchange with the environment along the actual phase-space trajectory and defines the instantaneous heat-like rate, 
$\dot Q^{*}$,
whereas the sum of the second and third terms accounts for the parametric energy input due to the protocols and defines the instantaneous work-like rate, 
$\dot W^{*}$.
Following the standard conventions of stochastic energetics \cite{talkner2020colloquium,jarzynski2004nonequilibrium},
their time integrals give
\begin{align}
\label{eq:work_time_int}
W^{*}(t_{\mathrm{eq}}|X_0) 
&=\int_{0}^{t_{\mathrm{eq}}}\mathrm dt\,
  \frac{\partial \mathcal H^{*}_{\beta}}{\partial\lambda}
  \,\dot\lambda(t)
  + \int_{0}^{t_{\mathrm{eq}}}\mathrm dt\,
  \frac{\partial \mathcal H^{*}_{\beta}}{\partial C}
  \,\dot C(t),
\\[4pt]
\label{eq:heat_time_int}
Q^{*}(t_{\mathrm{eq}}|X_0)
&=\int_{0}^{t_{\mathrm{eq}}}\mathrm dt\,
\bigl(\nabla_{X_{\mathcal S}}\mathcal H^{*}_{\beta}\bigr)
  \cdot\dot X_{\mathcal S}(t|X_0),
\end{align}
so that integrating Eq.~\eqref{eq:dH_dt} over 
$t\in[0,t_{\mathrm{eq}}]$ yields the total increment
\begin{equation}
\label{eq:I_total}
I^*(t_{\mathrm{eq}}|X_0)=W^{*}(t_{\mathrm{eq}}|X_0)+Q^{*}(t_{\mathrm{eq}}|X_0).
\end{equation}

\noindent
\textit{Term $II^*$.}
For fixed controls $\lambda(0)$ and $C(0)$, $II^*(t_{\mathrm{eq}}|X_0)$ can be written as
\begin{equation}
\label{eq:II_time_int}
II^*(t_{\mathrm{eq}}|X_0)
=\int_0^{t_{\mathrm{eq}}}\mathrm dt\;
\Bigl(\nabla_{X_{\mathcal S}}\mathcal H^{*}_{\beta}\bigl(X_{\mathcal S},\lambda(0),C(0)\bigr)\Bigr)
\cdot\dot X_{\mathcal S}.
\end{equation}
Here the gradient is evaluated with respect to 
$\mathcal H^{*}_{\beta}$ frozen at $\lambda(0)$ and $C(0)$,
whereas $\dot X_{\mathcal S}$
belongs to the driven trajectory generated by $\lambda(t)$ and $C(t)$.
Consequently, $II^*$ is a reference functional that projects the driven generalized velocity $\dot X_{\mathcal S}$ onto the generalized force field $\nabla_{X_{\mathcal S}}\mathcal H^{*}_{\beta}$ of the initial protocol.
Combining Eqs.~\eqref{eq:I_total} and \eqref{eq:II_time_int}
with Eq.~\eqref{eq:split_diff}, we obtain
\begin{align}
\label{eq:DeltaH_heatwork_final}
\Delta\mathcal H^{*}_{\beta}
\bigl(\mathcal T_{t_{\mathrm{eq}}}^{\mathcal S}(X_0)\bigr)
&=
W^{*}(t_{\mathrm{eq}}|X_0)
+Q^{*}(t_{\mathrm{eq}}|X_0)
\nonumber\\&-II^*(t_{\mathrm{eq}}|X_0).
\end{align}
Substituting Eq.~\eqref{eq:DeltaH_heatwork_final}
into Eqs.~\eqref{eq:traj_ratio}-\eqref{eq:traj_moment}
yields the compact heat-work-feedback-like representation of the endpoint equalities,
\begin{align}
\label{eq:heatwork_ratio_final}
e^{-\beta\Delta F_{\mathcal S}^{*}(\beta)}
&=\frac{
\Big\langle
e^{-\beta[\,W^{*}+Q^{*}-II^*\,]}
\Big\rangle_{X_0}}
{1+\chi^2\bigl(P^{\mathcal S}_{t_{\mathrm{eq}}}\parallel P^{\mathcal S}_0\bigr)},
\\[4pt]
\label{eq:heatwork_moment_final}
e^{+\beta\Delta F_{\mathcal S}^{*}(\beta)}
&=\Big\langle
e^{+\beta[\,W^{*}+Q^{*}-II^*\,]}
\Big\rangle_{X_0}.
\end{align}
The decomposition isolates three pathwise contributions:  
(i) $W^{*}$, mechanical work-like due to the applications of protocols;  
(ii) $Q^{*}$, heat-like exchanged during the whole process; and (iii) $II^*$, a {reference} (projection) functional obtained by contracting the driven velocity ($\dot X_{\mathcal S}$) with the $\lambda(0)$ and $C(0)$ force field ($\nabla_{X_{\mathcal S}}\mathcal H^{*}_{\beta}$).
\end{corollary}

\section{Frozen-driving
(coupling) view}
\label{Frozen-driving
(coupling) view}
\noindent
Although Eqs.~\eqref{eq:endpoint_ratio}, \eqref{eq:endpoint_moment}, and their trajectory counterparts \eqref{eq:traj_ratio}-\eqref{eq:traj_moment} remain valid for arbitrary simultaneous modulation of $\lambda(t)$ and $C(t)$, their full significance becomes clear once we examine them through two natural protocol perspectives that isolate which physical operation drives the system between the two equilibrium endpoints.
A first perspective is the frozen-coupling (driving) view, in which $C(t)\equiv C(0)$ is held fixed while $\lambda(t)$ is varied, as in Jarzynski-type protocols.
A second, complementary perspective is the frozen-driving (coupling) view, in which $\lambda(t)\equiv\lambda(0)$ is fixed and the interaction $C(t)$ is changed (see Fig.~\ref{fig1}), mirroring the structure of conventional free energy perturbation (FEP).
When the endpoint equalities are interpreted through these lenses, their thermodynamic content becomes more transparent. 
In the present work, we adopt the frozen-driving viewpoint to analyze interaction-changing protocols and to clarify their connection to coupling-based free energy estimators such as FEP.
The companion Letter develops the complementary frozen-coupling viewpoint, showing how the same framework recovers Jarzynski-type relations in the appropriate limit and extends them to strong system-environment coupling even when the dynamics violate Liouvillian or detailed-balance constraints.
It is important to note that both Jarzynski’s equality and FEP formulas inherit the protocol-agnostic character.
However, their standard derivations and applications are organized around these two aforementioned scenarios.
\noindent
\begin{corollary}[Frozen driving: endpoint forms in terms of system-environment interaction]
\label{cor:fixedlambda_endpoint}
For a fixed driving protocol $\lambda(t)\equiv \lambda(0)$, the HMF shift between the two equilibrium endpoints becomes
\begin{widetext}
\begin{align}
\Delta \mathcal H^{*}_{\beta}(X_{\mathcal S})=
\mathcal H^{*}_{\beta}\bigl(X_{\mathcal S},\lambda(0),C(t_{\mathrm{eq}})\bigr)
- \mathcal H^{*}_{\beta}\bigl(X_{\mathcal S},\lambda(0),C(0)\bigr)
\label{eq:DeltaHstar_endpoint_fixed}=
-\frac{1}{\beta}
\ln\frac{\mathcal M(X_{\mathcal S},C(t_{\mathrm{eq}}),\beta)}
{\mathcal M(X_{\mathcal S},C(0),\beta)} .
\end{align}
\end{widetext}
The cancellation of the bare system Hamiltonian in Eq.~\eqref{eq:HMF_def} leaves only the environment-interaction contribution $\mathcal M$.  
Inserting Eq.~\eqref{eq:DeltaHstar_endpoint_fixed} into Eqs.~\eqref{eq:endpoint_ratio}-\eqref{eq:endpoint_moment} and using the average in Eq.~\eqref{eq:avg_S} gives
\begin{align}
e^{-\beta \Delta F^{*}_{\mathcal S}(\beta)}
&=
\frac{
\Big\langle
\dfrac{\mathcal M(X_{\mathcal S},C(t_{\mathrm{eq}}),\beta)}
      {\mathcal M(X_{\mathcal S},C(0),\beta)}
\Big\rangle_{\mathcal S}
}{
1+\chi^2(P^{\mathcal S}_{t_{\mathrm{eq}}}\parallel P^{\mathcal S}_{0})
},
\label{eq:fixedlambda_endpoint_ratio}
\end{align}
and the complementary form
\begin{align}
e^{+\beta \Delta F^{*}_{\mathcal S}(\beta)}
&=
\Big\langle
\dfrac{\mathcal M(X_{\mathcal S},C(0),\beta)}
      {\mathcal M(X_{\mathcal S},C(t_{\mathrm{eq}}),\beta)}
\Big\rangle_{\mathcal S}.
\label{eq:fixedlambda_endpoint_moment}
\end{align}
In this regime, the free energy change is governed solely by ratios of the environment-interaction functional $\mathcal M$ evaluated at $C(0)$ and $C(t_{\mathrm{eq}})$.
\end{corollary}
\begin{corollary}[Frozen driving: trajectory forms in terms of system-environment interaction]
\label{cor:fixedlambda}
When $\lambda(t)\equiv \lambda(0)$, the trajectory version of the HMF increment reads
\begin{align}
\Delta\mathcal H^{*}_{\beta}\bigl(\mathcal T_{t_{\mathrm{eq}}}^{\mathcal S}(X_0)\bigr)
&=-\frac{1}{\beta}\ln\frac{\mathcal M(\mathcal T_{t_{\mathrm{eq}}}^{\mathcal S}(X_0),C(t_{\mathrm{eq}}),\beta)
}{
\mathcal M(\mathcal T_{t_{\mathrm{eq}}}^{\mathcal S}(X_0),C(0),\beta)
}.
\label{eq:DeltaHstar_traj_fixed}
\end{align}
Substituting Eq.~\eqref{eq:DeltaHstar_traj_fixed} into the trajectory equalities \eqref{eq:traj_ratio}-\eqref{eq:traj_moment} yields
\begin{align}
e^{-\beta \Delta F^{*}_{\mathcal S}(\beta)}
&=
\frac{
\Big\langle
\dfrac{
\mathcal M(\mathcal T_{t_{\mathrm{eq}}}^{\mathcal S}(X_0),C(t_{\mathrm{eq}}),\beta)
}{
\mathcal M(\mathcal T_{t_{\mathrm{eq}}}^{\mathcal S}(X_0),C(0),\beta)
}
\Big\rangle_{X_0}
}{
1+\chi^2(P^{\mathcal S}_{t_{\mathrm{eq}}}\parallel P^{\mathcal S}_0)
}
\label{eq:fixedlambda_ratio}
\end{align}
and
\begin{align}
e^{+\beta \Delta F^{*}_{\mathcal S}(\beta)}
&=
\Big\langle
\dfrac{
\mathcal M(\mathcal T_{t_{\mathrm{eq}}}^{\mathcal S}(X_0),C(0),\beta)
}{
\mathcal M(\mathcal T_{t_{\mathrm{eq}}}^{\mathcal S}(X_0),C(t_{\mathrm{eq}}),\beta)
}
\Big\rangle_{X_0}.
\label{eq:fixedlambda_moment}
\end{align}
If $\lambda(t)$ becomes time dependent, the simplification in Eqs.~\eqref{eq:DeltaHstar_endpoint_fixed} and \eqref{eq:DeltaHstar_traj_fixed} no longer holds and the full HMF structure must be used.
\end{corollary}
\section{Underlying identity and FEP}
\label{Underlying identity and FEP}
\noindent
The central identity underlying all our derivations is
\begin{equation}
\label{eq:central_identity}
e^{-\beta \Delta F^{*}_{\mathcal S}(\beta)}
=
\frac{\mathcal Z^{*}_{\mathcal S}\bigl(
\lambda(t_{\mathrm{eq}}),C(t_{\mathrm{eq}}),\beta
\bigr)}
{\mathcal Z^{*}_{\mathcal S}\bigl(
\lambda(0),C(0),\beta
\bigr)},
\end{equation}
where $\mathcal Z^{*}_{\mathcal S}(\lambda,C,\beta)$ is the HMF partition function defined by Eq.~\eqref{eq:Z_system}.  
By construction, the composite partition function factorizes as \cite{talkner2020colloquium}
\begin{equation}
\label{eq:factorization}
\mathcal Z_{\mathcal S+\mathcal E}(\lambda,C,\beta)
=
\mathcal Z^{*}_{\mathcal S}(\lambda,C,\beta)\,
\mathcal Z_{\mathcal E}(\beta),
\end{equation}
where $\mathcal Z_{\mathcal E}(\beta)$ depends only on the temperature.  
Since $\mathcal Z_{\mathcal E}(\beta)$ cancels in ratios, Eq.~\eqref{eq:central_identity} is equivalent to the composite equilibrium ratio
\begin{align}
\label{eq:ratio_composite}
\frac{\mathcal Z^{*}_{\mathcal S}\bigl(
\lambda(t_{\mathrm{eq}}),C(t_{\mathrm{eq}}),\beta
\bigr)}
{\mathcal Z^{*}_{\mathcal S}\bigl(
\lambda(0),C(0),\beta
\bigr)}
=
\frac{\mathcal Z_{\mathcal S+\mathcal E}\bigl(
\lambda(t_{\mathrm{eq}}),C(t_{\mathrm{eq}}),\beta
\bigr)}
{\mathcal Z_{\mathcal S+\mathcal E}\bigl(
\lambda(0),C(0),\beta
\bigr)}.
\end{align} 
Writing Eq.~\eqref{eq:ratio_composite} in terms of composite free energies yields
\begin{equation}
\label{eq:central_free_energy_identity}
e^{-\beta \Delta F^{*}_{\mathcal S}(\beta)}
=
e^{-\beta \Delta F_{\mathcal S+\mathcal E}(\beta)},
\end{equation}
with
\begin{align}
\label{eq:DeltaF_composite}
\Delta F_{\mathcal S+\mathcal E}(\beta)
&=
F_{\mathcal S+\mathcal E}\bigl(
\lambda(t_{\mathrm{eq}}),C(t_{\mathrm{eq}}),\beta
\bigr)
\nonumber\\
&
-
F_{\mathcal S+\mathcal E}\bigl(
\lambda(0),C(0),\beta
\bigr).
\end{align}
Thus, the open-system free energy difference encoded by the HMF coincides with the free energy difference of the composite system ($\mathcal{S}+\mathcal{E}$) evaluated at the same pair of protocol endpoints $(\lambda(0),C(0))$ and $(\lambda(t_{\mathrm{eq}}),C(t_{\mathrm{eq}}))$.
In conventional FEP and most of its applications, the two endpoint equilibrium states of the composite $\mathcal S+\mathcal E$ are chosen such that the initial state is non-interacting, while the final state incorporates the full system-environment interaction \cite{chipot2007free,zwanzig1954high}.  
To represent this standard situation, we consider an interaction potential of the form
\begin{equation}
\mathcal V_{\mathcal S\mathcal E}\bigl(X_{\mathcal S},X_{\mathcal E},C(t)\bigr)
=
C(t)\,\mathcal U_{\mathcal S\mathcal E}\bigl(X_{\mathcal S},X_{\mathcal E}\bigr),
\end{equation}
so that the coupling protocol $C(t)$ linearly controls a fixed interaction channel $\mathcal U_{\mathcal S\mathcal E}(X_{\mathcal S},X_{\mathcal E})$. 
Although the FEP derivation is formally agnostic with respect to the precise choice of protocol \cite{chipot2007free}, it is usually presented in the frozen-driving viewpoint, where
\begin{equation}
\label{eq:FEP_choice}
(\lambda(0),C(0))=(\lambda_0,0),
\qquad
(\lambda(t_{\mathrm{eq}}),C(t_{\mathrm{eq}}))=(\lambda_0,C(t_{\mathrm{eq}})),
\end{equation}
so that the driving protocol $\lambda(t)$ is held fixed at $\lambda_0$ and only the coupling protocol is changed from $0$ to $C(t_{\mathrm{eq}})$ such that the interaction energy $\mathcal{V}_\mathcal{SE}$ is zero at $t=0$.  
The resulting free energy difference is then expressed as an exponential average over the reference (non-interacting) ensemble \cite{zwanzig1954high}. 
We now derive this relation explicitly within our notation and connect it to Eq.~\eqref{eq:central_free_energy_identity}. For the endpoint choice \eqref{eq:FEP_choice}, the composite Hamiltonians read
\begin{align}
\label{eq:FEP_H0}
\mathcal H_0(X)
&\equiv
\mathcal H_{\mathcal S+\mathcal E}\bigl(X,\lambda_0,C(0)=0\bigr)
\nonumber\\
&=
\mathcal H_{\mathcal S}(X_{\mathcal S},\lambda_0)
+
\mathcal H_{\mathcal E}(X_{\mathcal E})+ \underbrace{C(0)\,\mathcal U_{\mathcal S\mathcal E}(X_{\mathcal S},X_{\mathcal E})}_{0}
\\[4pt]
\label{eq:FEP_H1}
\mathcal H_1(X)
&\equiv
\mathcal H_{\mathcal S+\mathcal E}\bigl(X,\lambda_0,C(t_{\mathrm{eq}})\bigr)
\nonumber\\
&=
\mathcal H_{\mathcal S}(X_{\mathcal S},\lambda_0)
+
\mathcal H_{\mathcal E}(X_{\mathcal E})
+
\underbrace{C(t_{\mathrm{eq}})\,\mathcal U_{\mathcal S\mathcal E}(X_{\mathcal S},X_{\mathcal E})}_{\mathcal V_{\mathcal S\mathcal E}\bigl(X_{\mathcal S},X_{\mathcal E},C(t_{\mathrm{eq}})\bigr)},
\end{align}
so that their difference is purely given by the interaction,
\begin{align}
\label{eq:FEP_H_diff}
\mathcal H_1(X)-\mathcal H_0(X)
= \mathcal V_{\mathcal S\mathcal E}\bigl(X_{\mathcal S},X_{\mathcal E},C(t_{\mathrm{eq}})\bigr).
\end{align}
The associated composite partition functions are
\begin{align}
\label{eq:FEP_Z0}
\mathcal Z_0
&\equiv
\mathcal Z_{\mathcal S+\mathcal E}(\lambda_0,0,\beta)
=
\int \mathrm dX\; e^{-\beta \mathcal H_0(X)},
\\[4pt]
\label{eq:FEP_Z1}
\mathcal Z_1
&\equiv
\mathcal Z_{\mathcal S+\mathcal E}(\lambda_0,1,\beta)
=
\int \mathrm dX\; e^{-\beta \mathcal H_1(X)},
\end{align}
and define the initial and final composite free energies,
\begin{align}
\label{eq:FEP_F0}
F_0(\beta)
&\equiv
F_{\mathcal S+\mathcal E}(\lambda_0,0,\beta)
=
-\,\frac{1}{\beta}\ln \mathcal Z_0,
\\[4pt]
\label{eq:FEP_F1}
F_1(\beta)
&\equiv
F_{\mathcal S+\mathcal E}(\lambda_0,1,\beta)
=
-\,\frac{1}{\beta}\ln \mathcal Z_1.
\end{align}
The composite free energy difference is
\begin{equation}
\label{eq:FEP_DeltaF_SE}
\Delta F_{\mathcal S+\mathcal E}(\beta)
=
F_1(\beta)-F_0(\beta)
=
-\,\frac{1}{\beta}\ln\frac{\mathcal Z_1}{\mathcal Z_0}.
\end{equation}
To obtain the FEP identity, we now express the ratio $\mathcal Z_1/\mathcal Z_0$ as an average over the initial non-interacting equilibrium state.  
Starting from the definition of $\mathcal Z_1$ and inserting the identity
$1 = e^{-\beta \mathcal H_0(X)} / e^{-\beta \mathcal H_0(X)}$ inside the integral, we find
\begin{align}
\label{eq:FEP_Z_ratio_step}
\frac{\mathcal Z_1}{\mathcal Z_0}
&=
\frac{1}{\mathcal Z_0}
\int \mathrm dX\; e^{-\beta \mathcal H_1(X)}
\nonumber\\
&=
\int \mathrm dX\;
e^{-\beta[\mathcal H_1(X)-\mathcal H_0(X)]}\,
\frac{e^{-\beta \mathcal H_0(X)}}{\mathcal Z_0}
\nonumber\\
&=
\int \mathrm dX\;
e^{-\beta[\mathcal H_1(X)-\mathcal H_0(X)]}\,
P_0(X,\beta),
\end{align}
where $P_0(X,\beta)$ is the canonical equilibrium distribution of the uncoupled composite
\begin{equation}
\label{eq:FEP_P0}
P_0(X,\beta)
=
\frac{e^{-\beta \mathcal H_0(X)}}{\mathcal Z_0}.
\end{equation}
Using Eq.~\eqref{eq:FEP_H_diff} we obtain
\begin{align}
\label{eq:FEP_Z_ratio_average}
\frac{\mathcal Z_1}{\mathcal Z_0}
&=
\int \mathrm dX\;
e^{-\beta\,\mathcal V_{\mathcal S\mathcal E}(X_{\mathcal S},X_{\mathcal E}, C(t_{\mathrm{eq}}))}\,
P_0(X,\beta)
\nonumber\\
&=
\Big\langle
e^{-\beta\,\mathcal V_{\mathcal S\mathcal E}(X_{\mathcal S},X_{\mathcal E}, C(t_{\mathrm{eq}}))}
\Big\rangle_{0},
\end{align}
where
\begin{equation}
\label{eq:FEP_avg_def}
\langle\bullet\rangle_{0}
\equiv
\int \mathrm dX\; \bullet\, P_0(X,\beta)
\end{equation}
denotes an average over the non-interacting reference ensemble.  
Combining Eqs.~\eqref{eq:FEP_DeltaF_SE} and \eqref{eq:FEP_Z_ratio_average} leads to the standard Zwanzig FEP identity \cite{chipot2007free},
\begin{equation}
\label{eq:FEP_final_identity}
e^{-\beta \Delta F_{\mathcal S+\mathcal E}(\beta)}
=
\Big\langle
e^{-\beta\,\mathcal V_{\mathcal S\mathcal E}(X_{\mathcal S},X_{\mathcal E}, C(t_{\mathrm{eq}}))}
\Big\rangle_{0}.
\end{equation}
Using the equivalence \eqref{eq:central_free_energy_identity}, this can be written in the HMF language as
\begin{equation}
\label{eq:FEP_final_HMF}
e^{-\beta \Delta F^{*}_{\mathcal S}(\beta)}
=
\Big\langle
e^{-\beta\,\mathcal V_{\mathcal S\mathcal E}(X_{\mathcal S},X_{\mathcal E}, C(t_{\mathrm{eq}}))}
\Big\rangle_{0}.
\end{equation}

\section{Beyond FEP}
\label{Beyond FEP}
\noindent
Conventional FEP [Eq.~\eqref{eq:FEP_final_HMF}] is recovered in our framework as the frozen-driving realization of Eqs.~\eqref{eq:endpoint_ratio} and \eqref{eq:traj_ratio}, obtained by fixing $\lambda(t)\equiv \lambda_0$, choosing a non-interacting reference state $\mathcal V_{\mathcal S\mathcal E}(X_{\mathcal S},X_{\mathcal E},C(0))=0$, and interpreting the coupling change as an instantaneous quench \cite{davoudi2025work}.
In this setting the interaction is switched from zero to its final value while the driving parameter remains fixed.  
Within the unified structure of our endpoint and trajectory identities, the usual FEP construction represents just one limiting protocol among many admissible ways of connecting equilibrium endpoints.
In this wider setting, the formalism extends naturally beyond the traditional domain of FEP.
To make this connection explicit, we first revisit the environment functional $\mathcal M$ appearing in the HMF definition \eqref{eq:HMF_def},
\begin{equation}
\mathcal M(X_{\mathcal S},C,\beta)
=
\frac{1}{\mathcal Z_{\mathcal E}}
\int \mathrm dX_{\mathcal E}\,
e^{-\beta\,[\mathcal H_{\mathcal E}(X_{\mathcal E})
+\mathcal V_{\mathcal S\mathcal E}(X_{\mathcal S},X_{\mathcal E},C)]}.
\end{equation}
For the conventional FEP setting, the interaction is of the form
$\mathcal V_{\mathcal S\mathcal E}(X_{\mathcal S},X_{\mathcal E},C(t))
= C(t)\,\mathcal U_{\mathcal S\mathcal E}(X_{\mathcal S},X_{\mathcal E})$
with $C(0)=0$.  
At $t=0$ the composite is uncoupled, $\mathcal V_{\mathcal S\mathcal E}=0$, so the integral reproduces the environment partition function.
Consequently,
\begin{equation}
\mathcal M(X_{\mathcal S},C(0)=0,\beta)
=
\frac{1}{\mathcal Z_{\mathcal E}}
\int \mathrm dX_{\mathcal E}\,
e^{-\beta\mathcal H_{\mathcal E}(X_{\mathcal E})}
=
1.
\label{eq:M_zero}
\end{equation}
In the frozen-driving regime, Cor.~\ref{cor:fixedlambda_endpoint} gives the pair of endpoint relations
\begin{align}
e^{-\beta \Delta F^{*}_{\mathcal S}(\beta)}
&=
\frac{
\big\langle
\mathcal M(X_{\mathcal S},C(t_{\mathrm{eq}}),\beta)
\big\rangle_{\mathcal S}
}{
1+\chi^2\bigl(P^{\mathcal S}_{t_{\mathrm{eq}}}\parallel P^{\mathcal S}_{0}\bigr)
},
\label{eq:FEP_like_endpoint_minus}
\end{align}
and
\begin{align}
e^{+\beta \Delta F^{*}_{\mathcal S}(\beta)}
&=
\Big\langle
\mathcal M(X_{\mathcal S},C(t_{\mathrm{eq}}),\beta)^{-1}
\Big\rangle_{\mathcal S}.
\label{eq:FEP_like_endpoint_plus}
\end{align}
The frozen-driving trajectory identities in Cor.~\ref{cor:fixedlambda} simplify in the same way.  
The trajectory counterparts of Eqs.~\eqref{eq:FEP_like_endpoint_minus} and \eqref{eq:FEP_like_endpoint_plus} become
\begin{align}
e^{-\beta \Delta F^{*}_{\mathcal S}(\beta)}
&=
\frac{
\Big\langle
\mathcal M\bigl(\mathcal T_{t_{\mathrm{eq}}}^{\mathcal S}(X_0),C(t_{\mathrm{eq}}),\beta\bigr)
\Big\rangle_{X_0}
}{
1+\chi^2\bigl(P^{\mathcal S}_{t_{\mathrm{eq}}}\parallel P^{\mathcal S}_{0}\bigr)
},
\label{eq:FEP_like_traj_minus}
\\[4pt]
e^{+\beta \Delta F^{*}_{\mathcal S}(\beta)}
&=
\Big\langle
\mathcal M\bigl(\mathcal T_{t_{\mathrm{eq}}}^{\mathcal S}(X_0),C(t_{\mathrm{eq}}),\beta\bigr)^{-1}
\Big\rangle_{X_0},
\label{eq:FEP_like_traj_plus}
\end{align}
where the averages are taken over initial conditions 
$X_0\sim P_0(X,\beta)$. Eqs.~\eqref{eq:FEP_like_traj_minus} and \eqref{eq:FEP_like_traj_plus} therefore provide a trajectory-based generalization of FEP.  
The interaction enters through $\mathcal M$ evaluated at the dynamically reached endpoint configurations $\mathcal T_{t_{\mathrm{eq}}}^{\mathcal S}(X_0)$, while the overlap correction retains its static definition via the chi-squared divergence of the endpoint marginals.  
Embedding the standard non-interacting reference into this structure clarifies in which precise sense our framework goes beyond conventional FEP.  
First, the role of phase-space overlap, which in standard FEP is not present and a serious numerical concern, appears here as an explicit multiplicative factor involving $\chi^2(P^{\mathcal S}_{t_{\mathrm{eq}}}\parallel P^{\mathcal S}_{0})$ that directly diagnoses when estimators are expected to be unreliable.  
Second, the free energy difference can be expressed as an average over trajectories underlying non-equilibrium processes for a given arbitrary dynamics and
coupling.  
Taken together, Eqs.~\eqref{eq:FEP_like_endpoint_minus}-\eqref{eq:FEP_like_traj_plus} extend FEP from the traditional instantaneous switching picture to general coupling protocols and dynamics, while keeping phase-space overlap and strong-coupling effects explicitly visible within a single unified formulation.
\begin{figure*}[t] 
  \centering
\includegraphics[width=\textwidth,height=0.4\textheight]{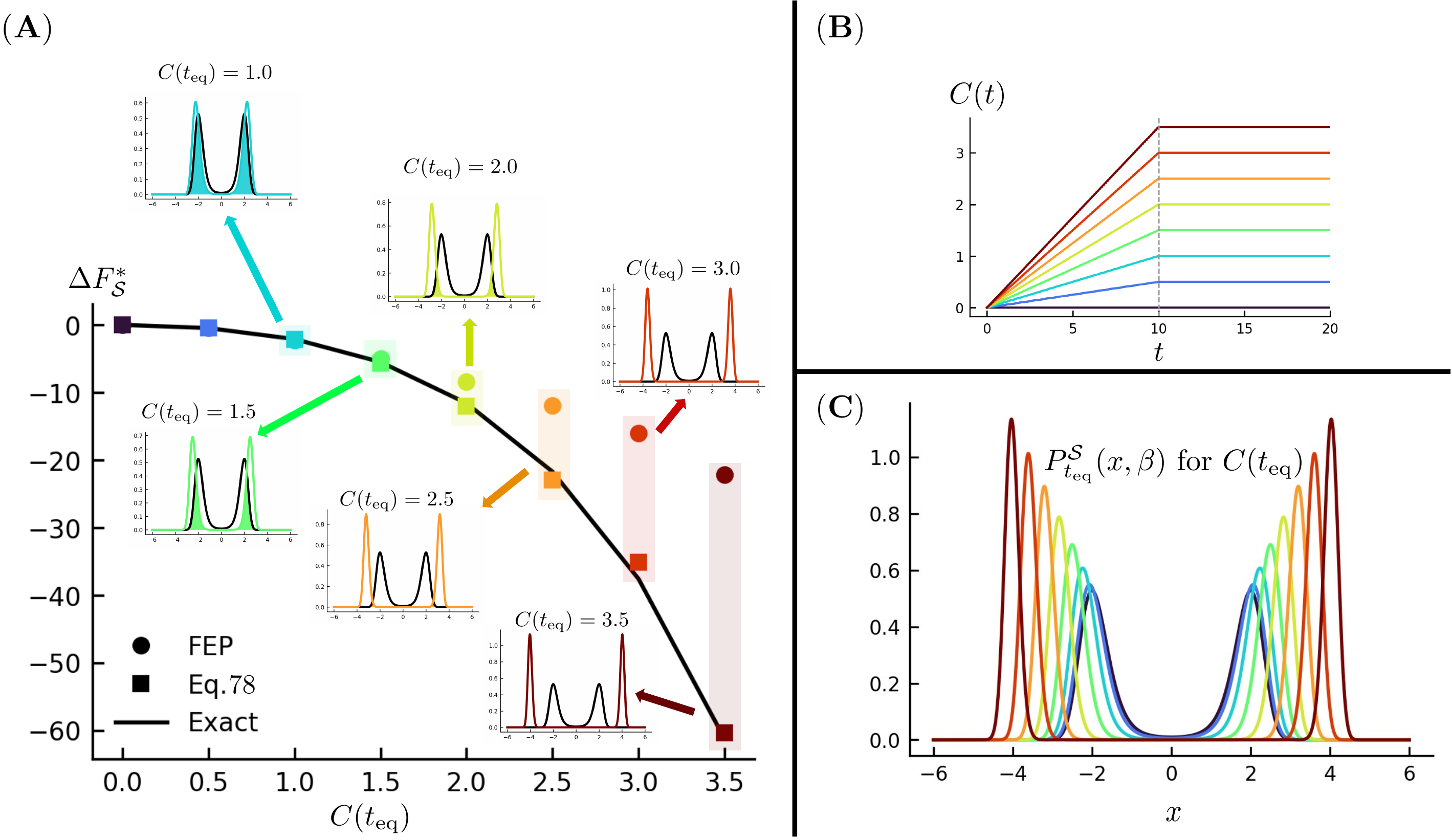}
  \caption{
\textbf{(A)} Open-system free energy differences 
$\Delta F^{*}_{\mathcal S}(\beta)$ as a function of the final coupling 
$C(t_{\mathrm{eq}})$.
The solid line denotes the exact HMF reference computed from 
Eq.~\eqref{eq:central_identity}. 
Circles show the conventional Zwanzig FEP estimator [Eq.~\eqref{eq:FEP_final_HMF}]. 
Squares show the trajectory counterpart of FEP derived in this work [Eq.~\eqref{eq:FEP_like_traj_minus}].
The small insets display the endpoint system marginals 
$P^{\mathcal S}_{t_{\mathrm{eq}}}(x,\beta)$ for representative coupling values, 
illustrating how increasing $C(t_{\mathrm{eq}})$  affects probabilities overlap.
\textbf{(B)} Coupling protocol $C(t)$ used in the frozen-driving regime: a linear ramp 
from $C(0)=0$ to $C(t_{\mathrm{eq}})$ followed by a relaxation stage, as in 
Eqs.~\eqref{eq:C_ramp_model} and \eqref{eq:C_relax_model}. 
\textbf{(C)} Endpoint system marginals $P^{\mathcal S}_{t_{\mathrm{eq}}}(x,\beta)$ for each 
final couplings.
Here we demonstrate that, while conventional FEP becomes 
inaccurate at moderate and strong coupling due to poor overlap, the 
trajectory equality Eq.~\eqref{eq:FEP_like_traj_minus} remains accurate across the entire range.}
  \label{fig2}
\end{figure*}

\section{Validation}
\label{sec:validation_fep_like}
\noindent
We now validate the trajectory equality in Eq.~\eqref{eq:FEP_like_traj_minus} (see Fig.~\ref{fig2}). 
The test is organized in the frozen-driving regime of Sec.~\ref{Frozen-driving (coupling) view}, with a non-interacting reference state as in Eq.~\eqref{eq:FEP_choice}. 
This setup allows a direct comparison between the Zwanzig identity \eqref{eq:FEP_final_identity} and its trajectory counterpart \eqref{eq:FEP_like_traj_minus}.

\noindent
\textit{Model specification}.  
We consider a one-dimensional system $\mathcal S$ in a quartic double-well potential \cite{whitelam2025improving,hanggi1990reaction,li2021equilibrium,saito1976relaxation,sun2003equilibrium} coupled linearly to a harmonic environment $\mathcal E$. 
The bare Hamiltonians are
\begin{equation}
\mathcal H_{\mathcal S}(x,p_x;\lambda)
=
\frac{p_x^2}{2m}
+
U_{\mathcal S}(x;\lambda),
\label{eq:H_S_model}
\end{equation}
with
\begin{equation}
U_{\mathcal S}(x;\lambda)
=
\frac{1}{4}\bigl(x^2-\lambda\bigr)^2
\label{eq:U_S_model}
\end{equation}
and
\begin{equation}
\mathcal H_{\mathcal E}(y,p_y)
=
\frac{p_y^2}{2m}
+
\frac{1}{2}\omega^2 y^2.
\label{eq:H_E_model}
\end{equation}
The interaction channel is
\begin{equation}
\mathcal V_{\mathcal S\mathcal E}(x,y;C)
=
C\,x\,y.
\label{eq:V_SE_model}
\end{equation}
The total Hamiltonian follows Def.~\ref{def:fullH}. and we choose reduced units with $m=1$. 

\noindent
\textit{Derivation of the HMF and system marginal distribution}.  
We now derive the explicit HMF for the system.
For fixed $(\lambda_0,C)$, the full Hamiltonian reads
\begin{align}
\mathcal H_{\mathcal S+\mathcal E}(x,p_x,y,p_y;\lambda_0,C)
&\nonumber=
\mathcal H_{\mathcal S}(x,p_x;\lambda_0)
+
\mathcal H_{\mathcal E}(y,p_y)
\\&+
\mathcal V_{\mathcal S\mathcal E}(x,y;C),
\label{eq:HMF_DW_fullH}
\end{align} 
with $\mathcal H_{\mathcal S}$, $\mathcal H_{\mathcal E}$ and $\mathcal V_{\mathcal S\mathcal E}$ given by Eqs.~\eqref{eq:H_S_model}-\eqref{eq:V_SE_model}. 
According to the HMF definition in Eq.~\eqref{eq:HMF_def} and restricting to the configurational part (the momentum factors are independent of $x$ and can be absorbed into $\mathcal Z^*_{\mathcal S}$), we can write
\begin{align}
&\nonumber e^{-\beta \mathcal H^*_\beta(x;\lambda_0,C)} =
e^{-\beta U_{\mathcal S}(x;\lambda_0)}
\\&\times\underbrace{\frac{1}{\mathcal Z_{\mathcal E}(\beta)}
\int \mathrm d y\,\mathrm d p_y\,
\exp\Bigl(
-\beta\bigl(
\tfrac{p_y^2}{2}
+
\tfrac{1}{2}\omega^2 y^2
+
Cxy
\bigr)
\Bigr)}_{A}.
\label{eq:HMF_DW_def_Us_factor}
\end{align}  
\noindent
The environment partition function appears in the numerator of Eq.~\eqref{eq:HMF_DW_def_Us_factor} factorizes into momentum and configuration contributions, as in Eq.~\eqref{eq:Z_env},
\begin{equation}
\mathcal Z_{\mathcal E}(\beta)
=
\int\mathrm d p_y\,e^{-\beta p_y^2/2}
\int\mathrm d y\,e^{-\beta \omega^2 y^2/2}.
\label{eq:HMF_DW_ZE_factor}
\end{equation} 
The $p_y$ integral cancels between numerator and denominator.
Introducing the configurational environment partition function
\begin{equation}
Z_{\mathcal E}^{(y)}(\beta)
=
\int \mathrm d y\,\exp\Bigl(-\frac{\beta\omega^2 y^2}{2}\Bigr),
\label{eq:HMF_DW_ZEy_def}
\end{equation}
we can write
\begin{equation}
A
=
\frac{1}{Z_{\mathcal E}^{(y)}(\beta)}
\int\mathrm d y\,
\exp\Bigl(-\beta\Bigl(\frac{1}{2}\omega^2 y^2 + Cxy\Bigr)\Bigr).
\label{eq:HMF_DW_y_only}
\end{equation}
\noindent
To evaluate Eq.~\eqref{eq:HMF_DW_y_only}, we complete the square in $y$. 
The quadratic form in the exponent can be written as
\begin{align}
\frac{1}{2}\omega^2 y^2 + Cxy
&\nonumber=
\frac{1}{2}\omega^2\Bigl(
y^2 + 2\frac{C x}{\omega^2}y
\Bigr)
\\&=
\frac{1}{2}\omega^2\Bigl((y+a)^2 - a^2\Bigr),
\label{eq:HMF_DW_complete_square_1}
\end{align} 
where $a\equiv\frac{C x}{\omega^2}$.
Hence, Eq.~\eqref{eq:HMF_DW_y_only} gives
\begin{align}
A=
\exp\Bigl(\frac{\beta C^2 x^2}{2\omega^2}\Bigr)
\frac{1}{Z_{\mathcal E}^{(y)}(\beta)}
\int\mathrm d y\,
\exp\Bigl(-\beta\frac{1}{2}\omega^2(y+a)^2\Bigr).
\label{eq:HMF_DW_gaussian_shift}
\end{align} 
The integral in Eq.~\eqref{eq:HMF_DW_gaussian_shift} is invariant under the shift $y\mapsto y-a$ and therefore equals $Z_{\mathcal E}^{(y)}(\beta)$ defined in Eq.~\eqref{eq:HMF_DW_ZEy_def}. 
We thus obtain
\begin{equation}
A=
\exp\Bigl(\frac{\beta C^2 x^2}{2\omega^2}\Bigr).
\label{eq:HMF_DW_env_ratio_final}
\end{equation}
Inserting Eq.~\eqref{eq:HMF_DW_env_ratio_final} into Eq.~\eqref{eq:HMF_DW_def_Us_factor} yields
\begin{align}
e^{-\beta \mathcal H^*_\beta(x;\lambda_0,C)}=
\exp\Bigl(-\beta\Bigl(U_{\mathcal S}(x;\lambda_0)
-
\frac{C^2 x^2}{2\omega^2}\Bigr)\Bigr).
\label{eq:HMF_DW_exponent_final}
\end{align}
Taking the logarithm gives the HMF
\begin{equation}
\mathcal H^*_\beta(x;\lambda_0,C)
=
U_{\mathcal S}(x;\lambda_0)
-
\frac{C^2 x^2}{2\omega^2},
\label{eq:HMF_double_well_model}
\end{equation}
Inserting Eq.~\eqref{eq:HMF_double_well_model} into Eq.~\eqref{eq:Z_system} and Eq.~\eqref{eq:system_canonical} provides the  marginals $P_0^{\mathcal S}$ and $P_{t_{\mathrm{eq}}}^{\mathcal S}$ at the endpoints $(\lambda(0),C(0))$ and $(\lambda(t_{\mathrm{eq}}),C(t_{\mathrm{eq}}))$.

\noindent
\textit{Frozen-driving protocol and dynamics}.  
We adopt the frozen-driving viewpoint of Sec.~\ref{Frozen-driving (coupling) view} by fixing
\begin{equation}
\lambda(t)\equiv\lambda_0,
\qquad
C(0)=0,
\qquad
C(t_{\mathrm{eq}}),
\label{eq:frozen_driving_protocol_model}
\end{equation}
so that the system is transported between equilibrium states only by changing the coupling. 
During a ramp stage $0\le t\le t_{\mathrm{ramp}}$, the coupling increases linearly from $0$ to $C(t_{\mathrm{eq}})$,
\begin{equation}
C(t)
=
C(t_{\mathrm{eq}})\,\frac{t}{t_{\mathrm{ramp}}},
\qquad
0\le t\le t_{\mathrm{ramp}},
\label{eq:C_ramp_model}
\end{equation}
followed by a relaxation stage
\begin{equation}
C(t)\equiv C(t_{\mathrm{eq}}),
\qquad
t_{\mathrm{ramp}}\le t\le t_{\mathrm{eq}}.
\label{eq:C_relax_model}
\end{equation}
The resulting coupling schedule and relaxation window used in the simulations are illustrated in Fig.~\ref{fig2} (see panel \textbf{B}).
The composite evolves under overdamped Langevin dynamics \cite{seifert2012stochastic} consistent with the canonical measures in Eqs.~\eqref{eq:canonical_initial} and \eqref{eq:canonical_final}. 
Writing the total potential as
\begin{equation}
U_{\mathrm{tot}}(x,y;C)
=
U_{\mathcal S}(x;\lambda_0)
+
\frac{1}{2}\omega^2y^2
+
Cxy,
\label{eq:U_tot_model}
\end{equation}
the equations of motion are
\begin{equation}
\dot x
=
-\mu\,\partial_x U_{\mathrm{tot}}(x,y;C(t))
+
\sqrt{\frac{2\mu}{\beta}}\,\xi_x(t),
\label{eq:langevin_x_model}
\end{equation}
\begin{equation}
\dot y
=
-\mu\,\partial_y U_{\mathrm{tot}}(x,y;C(t))
+
\sqrt{\frac{2\mu}{\beta}}\,\xi_y(t),
\label{eq:langevin_y_model}
\end{equation}
where $\mu$ is a mobility \cite{spiechowicz2022diffusion} and $\xi_i(t)$ are independent unit white noises. 
For each fixed pair $(\lambda_0,C)$ this dynamics satisfies the fluctuation-dissipation relation and converges to the equilibrium state associated with $\mathcal H_{\mathcal S+\mathcal E}(x,y;\lambda_0,C)$. The corresponding endpoint system marginals for different final couplings are shown in Fig.~\ref{fig2} (see panel \textbf{C}). 
Choosing $t_{\mathrm{eq}}$ large compared to the relaxation time enforces the asymptotic equilibration condition in Eq.~\eqref{eq:system_relax}, so that the trajectory map $\mathcal T_{t_{\mathrm{eq}}}$ realizes the endpoint marginal $P^{\mathcal S}_{t_{\mathrm{eq}}}$ in Eq.~\eqref{eq:HMF_final}. 
Initial conditions $X_0=(x_0,y_0)$ are sampled from the uncoupled composite equilibrium in Eq.~\eqref{eq:canonical_initial} with $(\lambda(0),C(0))=(\lambda_0,0)$. 
At $C=0$ the composite density factorizes, so we draw
\begin{equation}
x_0\sim
\frac{1}{\mathcal Z_{\mathcal S}(\lambda_0,\beta)}
\exp\bigl(-\beta U_{\mathcal S}(x;\lambda_0)\bigr),
\label{eq:x0_sampling_model}
\end{equation}
and
\begin{equation}
y_0\sim
\frac{1}{\mathcal Z_{\mathcal E}(\beta)}
\exp\Bigl(-\frac{\beta\omega^2 y^2}{2}\Bigr),
\label{eq:y0_sampling_model}
\end{equation}
which realizes the non-interacting reference of Eq.~\eqref{eq:FEP_choice}, where the Zwanzig identity \eqref{eq:FEP_final_identity} applies.

\noindent
\textit{Estimators for $\Delta F^*_{\mathcal S}(\beta)$}.  
For each final coupling $C(t_{\mathrm{eq}})$ we compute the open-system free energy difference
\begin{equation}
\Delta F^*_{\mathcal S}(\beta)
=
F^*_{\mathcal S}(\lambda_0,C(t_{\mathrm{eq}}),\beta)
-
F^*_{\mathcal S}(\lambda_0,0,\beta),
\label{eq:DF_model}
\end{equation}
using three constructions.

\noindent
(i) \emph{Exact HMF reference}.  
Using the central identity in Eq.~\eqref{eq:central_identity}, we evaluate the HMF partition functions $\mathcal Z^*_{\mathcal S}(\lambda_0,C,\beta)$ by numerical quadrature of Eq.~\eqref{eq:Z_system} with the model HMF in Eq.~\eqref{eq:HMF_double_well_model}. 

\noindent
(ii) \emph{Zwanzig FEP}.  
We evaluate the interaction $\mathcal V_{\mathcal S\mathcal E}(X_{\mathcal S},X_{\mathcal E})$ on the uncoupled configurations $X_0\sim P_0(X,\beta)$ and estimate
\begin{equation}
\Big\langle
e^{-\beta \mathcal V_{\mathcal S\mathcal E}(X_{\mathcal S},X_{\mathcal E};C(t_{\mathrm{eq}}))}
\Big\rangle_0.
\label{eq:zwanzig_avg_model}
\end{equation}
This provides a Zwanzig estimator, Eq.~\eqref{eq:FEP_final_HMF}, for calculating the free energy differences.

\noindent
(iii) \emph{Trajectory counterpart of FEP}.  
The main test concerns the trajectory counterpart of FEP, Eq.~\eqref{eq:FEP_like_traj_minus}. 
We evaluate $\mathcal M$ at the endpoint positions $x_{\mathrm{eq}} = \mathcal T_{t_{\mathrm{eq}}}^{\mathcal S}(X_0)$ for each trajectory. Subsequently, we compute the sample average $\langle \mathcal M(\mathcal T_{t_{\mathrm{eq}}}^{\mathcal S}(X_0), C(t_{\mathrm{eq}}), \beta) \rangle_{X_0}$ and divide it by the overlap factor $1 + \chi^2(P^{\mathcal S}_{t_{\mathrm{eq}}} \parallel P^{\mathcal S}_{0})$, obtained from Eq.~\eqref{eq:chi2_def}.
The exact HMF reference, the Zwanzig estimator, and the trajectory-based estimator are compared as a function of the final coupling in Fig.~\ref{fig2} (see panel \textbf{A}). 
For completeness, the numerical values of all simulation parameters used in the validation of Eq.~\eqref{eq:FEP_like_traj_minus} are reported in Table~\ref{tab:params_fep_like}.

\begin{table}[h]
\centering
\caption{Numerical inputs for the validation runs (frozen driving; overdamped Langevin dynamics with linear coupling ramp). All quantities are in reduced, dimensionless units.}
\label{tab:params_fep_like}
\begin{ruledtabular}
\begin{tabular}{l l}
Quantity & Value \\
\hline
Temperature $\beta^{-1}/k_B$ & $1.0$ \\
Double–well control parameter $\lambda_0$ & $4.0$ \\
Environment frequency $\omega$ & $1.0$ \\
Mobility $\mu$ & $1.0$ \\
Initial coupling $C_0$ & $0.0$ \\
Ramp duration $t_{\mathrm{ramp}}$ & $10.0$ \\
Relaxation duration $t_{\mathrm{relax}}$ & $10.0$ \\
Total protocol time $t_{\mathrm{eq}}$ & $20.0$ \\
Time step $\Delta t$ & $10^{-2}$ \\
Trajectory ensemble size $N_{\mathrm{traj}}$ & $40000$ \\
FEP ensemble size $N_{\mathrm{fep}}$ & $40000$ \\
\end{tabular}
\end{ruledtabular}
\end{table}

\section{Conclusions and outlook}
\label{sec:conclusions}

\noindent
We have presented a general fluctuation framework for classical open systems that unifies equilibrium endpoint relations, nonequilibrium trajectory formulations, and strong-coupling thermodynamics within a single mathematical structure.
A central feature of this framework is the systematic use of the HMF, which provides the correct thermodynamic potential for a system that interacts with its environment at finite strength.
The necessity of using the HMF in place of the bare system Hamiltonian is a central theme in modern thermodynamics~\cite{talkner2020colloquium}, and the present work builds directly upon this principle. 
Our starting point is a fully general description of the composite $\mathcal S+\mathcal E$, formulated through an abstract trajectory map $\mathcal T_t$ that does not presuppose any constraints on dynamics.
This allows us to treat nonequilibrium evolution and coupling protocols without imposing microscopic reversibility (Liouvillian structure), detailed balance (DB), fluctuation-dissipation theorem (FDT) or local detailed balance (LDB) \cite{jarzynski1997equilibrium,jarzynski1997nonequilibrium,jarzynski2004nonequilibrium,sagawa2010generalized,speck2007jarzynski}. 
Within this setting, the canonical endpoints of the evolution are taken as bona fide equilibrium states of the composite, and all thermodynamic quantities are defined through the HMF.
Thm.~\ref{thm:endpoint_equalities} demonstrates that the open system free energy difference $\Delta F^{*}_{\mathcal S}(\beta)$ between arbitrary endpoint states admits exact representations in terms of the HMF shift and the chi-squared divergence between the endpoint system marginals. 
This chi-squared factor quantifies the statistical change in the system distribution induced by strong coupling, explicitly linking the free energy difference to the mismatch between endpoint ensembles. 
Such mismatches, and their impact on thermodynamic potentials and information measures, are widely emphasized in strong coupling formulations of open system thermodynamics~\cite{talkner2020colloquium,strasberg2020measurability} but typically appear only implicitly.
Here they enter at the level of exact fluctuation relations, providing a clean
and quantitative diagnostic. 
A notable feature of our derivation is that it does not rely on any particular dynamical generator.
Once the assumption of asymptotic equilibration is made, Thm.~\ref{thm:traj_endpoint} elevates the endpoint identities to exact trajectory relations for general nonequilibrium processes.
This result establishes that the free energy difference between two canonical states can be computed from nonequilibrium trajectories, even when strong coupling prevents the existence of a clear separation between system and environment energies.
Moreover, Cor.~\ref{cor:heatwork_fixedC} provides an exact decomposition of the
trajectory HMF increment into three pathwise functionals: (i) a work-like contribution originating from explicit changes in $(\lambda(t),C(t))$; (ii) a heat-like contribution representing energetic exchange with the environment along the actual trajectory; and (iii) a reference functional that projects the driven velocities onto the force field defined by the initial protocol. 
This decomposition offers a consistent strong coupling analogue of the first-law structure familiar from stochastic energetics and nonequilibrium work relations~\cite{jarzynski2004nonequilibrium}.
Refs.~\cite{talkner2020colloquium,strasberg2020measurability} have repeatedly emphasized the difficulty of defining heat, work, and internal energy in regimes where system-environment correlations play an essential role.
The representation obtained here provides one of the few fully general and exact constructions in which these quantities appear with clear operational meaning. 
\noindent
A major conceptual outcome of this work is the identification of two complementary protocol views---frozen coupling and frozen driving---that divide the fluctuation framework into two branches.
In the frozen-driving view analyzed here, the driving parameter $\lambda(t)$ is held constant while the interaction $C(t)$ transports the system between endpoints.
In this regime, the endpoint and trajectory identities simplify substantially, the bare system Hamiltonian cancels from the HMF, and the free energy difference does not depend on the system Hamiltonian directly but entirely on the environment functional $\mathcal M(X_{\mathcal S},C,\beta)$.
Corollaries~\ref{cor:fixedlambda_endpoint} and~\ref{cor:fixedlambda} show that all exponential relations can be expressed as ratios of $\mathcal M$ evaluated at $C(0)$ and $C(t_{\mathrm{eq}})$, together with the overlap factor.
This yields a trajectory-based generalization of Zwanzig's FEP formula [Eqs.~\eqref{eq:FEP_like_traj_minus}--\eqref{eq:FEP_like_traj_plus}], valid
under arbitrary dynamics and arbitrary strong coupling. 
Standard FEP emerges as the special case in which the initial composite is uncoupled and the interaction is applied instantaneously. 
The complementary "JE branch" is developed in a separate Letter~\cite{rahbar2025exact}, where the frozen-coupling view leads to a
generalized Jarzynski-type identity that remains exact at strong coupling and
does not require the dynamics to satisfy constraints such as microscopic reversibility, DB, FDT or LDB. This unification is consistent with long-standing observations in, e.g. Refs.~\cite{jarzynski1997equilibrium,jarzynski1997nonequilibrium,talkner2020colloquium} that FEP and JE represent boundary cases of a more
general, strong-coupling fluctuation theory. 
The numerical validation in Sec.~\ref{sec:validation_fep_like} reveals the practical benefits of the present framework.
For a strongly coupled system evolving under overdamped Langevin dynamics, we compared the three estimates of $\Delta F^{*}_{\mathcal S}(\beta)$ across a range of coupling strengths.
When the endpoint marginals exhibit substantial overlap, the classical Zwanzig estimator, the trajectory-based estimator, and the exact HMF reference agree.
At moderate and strong coupling, however, the distributions separate, the phase-space overlap deteriorates, and the classical FEP estimator fails.
In contrast, the trajectory-based estimator remains accurate across the full coupling range, precisely because it incorporates the overlap factor and relies on HMF-based energetics interactions.
These results demonstrate that the explicit chi-squared correction supplies a principled mechanism for detecting and correcting the primary failure mode of conventional FEP.
They also illustrate how strong-coupling corrections manifest operationally in a model system, complementing the conceptual discussions in the strong-coupling literature~\cite{talkner2020colloquium}.

The broader implications of this work span both numerical methodology and strong-coupling thermodynamics. From a computational perspective, the identities derived here suggest new ways to design and optimize free energy calculations in open systems, because their explicit dependence on the chi-squared divergence identifies, in a quantitative manner, how protocol choices affect the overlap between endpoint marginals.
This provides a natural conceptual bridge to long-standing strategies developed within the FEP community, where the reliability of a calculation is known to be controlled by the degree of ensemble overlap. 
In conventional FEP, overlap is often improved by inserting intermediate states and stratifying ensembles~\cite{torrie1977nonphysical,torrie1974monte,shirts2008statistically,wu2004asymmetric,reinhardt2020determining,matsunaga2022use}, or by applying enhanced sampling methods such as umbrella sampling, metadynamics, accelerated MD, and related biasing schemes~\cite{kastner2011umbrella,hamelberg2004accelerated,barducci2008well,plotnikov2014computing,kastner2005bridging,PhysRevLett.113.090601,bernardi2015enhanced,laio2008metadynamics}.
Building on this foundation, recent progress in alchemical free energy calculations and advanced enhanced-sampling schemes for free-energy landscapes~\cite{zhang2024alchemical,wade2022alchemical,knirsch2025practical,invernizzi2020unified,falkner2024enhanced} has made the notion of overlap even more explicit, with optimized coupling-schedules and variational biases designed to flatten barriers and improve phase-space mixing.
Adaptive schemes for constructing alchemical paths  aim to enhance or optimize phase-space overlap directly~\cite{zhang2024alchemical,knirsch2025practical}, often selecting intermediate states to stabilize estimators or reduce rare-event contributions. 
In parallel, thermodynamically consistent open-system coupling procedures based on adaptive-resolution and particle-domain methods~\cite{PhysRevLett.108.170602,PhysRevLett.129.230603} demonstrate that correct endpoint statistics and, in particular, matching the system marginal to a target ensemble are essential for reliable equilibrium and nonequilibrium characterizations.
In adaptive-resolution simulations \cite{PhysRevLett.108.170602}, a thermodynamic force or free energy compensation term is tuned so that density and structural properties in the atomistic region reproduce those of a fully atomistic reference, effectively enforcing the desired subsystem marginal.
In particle-continuum couplings with fluctuating hydrodynamics \cite{PhysRevLett.129.230603}, the exchange of matter and energy is constructed so that the particle domain exhibits the same thermodynamic fluctuations as a macroscopic reservoir, again using subsystem statistics as the standard of correctness.
Seen from this perspective, the present framework provides a unifying principle behind these algorithmic developments.
By identifying the chi-squared divergence $\chi^{2}\bigl(P^{\mathcal S}_{t_{\mathrm{eq}}}\Vert P^{\mathcal S}_{0}\bigr)$ as the exact measure of ensemble mismatch entering a strong-coupling fluctuation identity, it offers a direct and thermodynamically grounded criterion for choosing, optimizing, or adaptively refining intermediate states.
Because $\chi^{2}$ depends only on endpoint marginals, it becomes a practical diagnostic for assessing overlap and a principled target for adaptive schemes designed to maintain ensemble similarity along a chosen thermodynamic path. 
Another natural extension of this framework is its integration with machine-learned collective variables and unsupervised dimensionality-reduction methods.
Recent studies have shown that low-dimensional representations can be constructed to preserve or enhance overlap between ensembles~\cite{swinburne2018unsupervised,invernizzi2020unified,mehdi2024enhanced,fu2024collective,mendels2022collective}.
Within such approaches, the chi-squared divergence can provide a concrete objective for selecting or refining collective variables and for determining when enhanced sampling has reached sufficient coverage of the relevant configuration space. 
Similarly, multi-stage and replica-based strategies~\cite{torrie1977nonphysical,kastner2005bridging,kastner2011umbrella,plotnikov2014computing,wade2022alchemical,pregeljc2025efficient,hsu2024replica} can naturally be reinterpreted as sampling schemes that explore controlled sequences of effective Hamiltonians (or HMFs) along paths in $(\lambda,C,\beta)$. 
Rewriting these constructions in the HMF language can clarify when strong-coupling effects are absorbed into effective potentials and when they might be treated explicitly at the level of open-system energetics.
On the conceptual side, the framework contributes to a growing effort to establish a consistent thermodynamic theory for strongly coupled open systems, complementing modern analyses that highlight the ambiguity of heat, work, entropy production, and free energy for strongly coupled open system ~\cite{talkner2020colloquium,strasberg2020measurability}. Expressing observables in terms of the HMF and monitoring ensemble discrepancies through $\chi^{2}$ suggests a route by which such ambiguities can be reduced, while the trajectory-level heat–work–feedback decomposition gives an explicit operational interpretation of energetic contributions in nonequilibrium processes that sits naturally alongside earlier formulations~\cite{jarzynski2004nonequilibrium, talkner2020colloquium,sagawa2010generalized}. 
We expect that the formal results developed here, together with those presented in the companion Letter~\cite{rahbar2025exact} and our recent related work~\cite{rahbar2025probabilistic,rahbar2025thermodynamic}, will provide a foundation for new algorithmic developments in molecular simulation of free energy calculations and will support further theoretical and experimental advances in the thermodynamics of strongly coupled open systems.

\begin{acknowledgments} We gratefully acknowledge ﬁnancial support by the DFG
under Germany's Excellence Strategy EXC 2089/2-390776260
(e-conversion).
\end{acknowledgments}

\bibliography{apssamp}

\end{document}